\documentclass[%
 superscriptaddress,
 preprint,
 nofootinbib,
 amsmath,amssymb,
 aps,
 pra,
 longbibliography,
 a4paper
]{revtex4-1}
\pdfoutput=1
\usepackage[normalem]{ulem}
\usepackage{footnote}
\usepackage{graphicx}% Include figure files
\usepackage{dcolumn}% Align table columns on decimal point
\usepackage{bm}% bold math
\usepackage{slashed}
\usepackage{color}
\usepackage{url}
\usepackage{comment}
\usepackage{physics}
\usepackage[hidelinks]{hyperref}
\usepackage{mathtools}

\usepackage{tikz}
\usetikzlibrary{quantikz2}

\renewcommand{\ket}[1]{|#1\rangle}
\renewcommand{\bra}[1]{\langle #1|}

\newcommand{\tens}[2]{#1^{\otimes #2}}

\begin{document}

\preprint{TU-1244, KEK-QUP-2024-0024}

\preprint{MIT-CTP/5798}

\title{
Entanglement-enhanced AC magnetometry in the presence of Markovian noises
}

\author{Thanaporn Sichanugrist}
\email{thanaporn@hep-th.phys.s.u-tokyo.ac.jp}
\affiliation{Department of Physics, The University of Tokyo, Tokyo 113-0033, Japan}

\author{Hajime Fukuda}
\email[Corresponding author: ]{hfukuda@hep-th.phys.s.u-tokyo.ac.jp}
\affiliation{Department of Physics, The University of Tokyo, Tokyo 113-0033, Japan}

\author{Takeo Moroi}
\email{moroi@hep-th.phys.s.u-tokyo.ac.jp}
\affiliation{Department of Physics, The University of Tokyo, Tokyo 113-0033, Japan}
\affiliation{International Center for Quantum-field Measurement Systems for Studies of the Universe and Particles (QUP), High Energy Accelerator Research Organization (KEK), 1-1 Oho, Tsukuba, Ibaraki 305-0801, Japan}

\author{Kazunori Nakayama}
\affiliation{Department of Physics, Tohoku University, Sendai, Miyagi 980-8578, Japan}
\affiliation{International Center for Quantum-field Measurement Systems for Studies of the Universe and Particles (QUP), High Energy Accelerator Research Organization (KEK), 1-1 Oho, Tsukuba, Ibaraki 305-0801, Japan}

\author{So Chigusa}
\affiliation{Theoretical Physics Group, Lawrence Berkeley National Laboratory, Berkeley, CA 94720, USA}
\affiliation{Berkeley Center for Theoretical Physics, Department of Physics,
University of California, Berkeley, CA 94720, USA}
\affiliation{Center for Theoretical Physics, Massachusetts Institute of Technology, Cambridge, MA 02139, USA}

\author{Norikazu Mizuochi}
\affiliation{International Center for Quantum-field Measurement Systems for Studies of the Universe and Particles (QUP), High Energy Accelerator Research Organization (KEK), 1-1 Oho, Tsukuba, Ibaraki 305-0801, Japan}
\affiliation{Institute for Chemical Research, Kyoto University, Gokasho, Uji-city, Kyoto 611-0011, Japan}
\affiliation{Center for Spintronics Research Network, Kyoto University, Uji, Kyoto 611-0011, Japan}

\author{Masashi Hazumi}
\affiliation{International Center for Quantum-field Measurement Systems for Studies of the Universe and Particles (QUP), High Energy Accelerator Research Organization (KEK), 1-1 Oho, Tsukuba, Ibaraki 305-0801, Japan}
\affiliation{Institute of Particle and Nuclear Studies (IPNS), KEK, Tsukuba, Ibaraki 305-0801, Japan}
\affiliation{Japan Aerospace Exploration Agency (JAXA), Institute of Space and Astronautical Science (ISAS), Sagamihara, Kanagawa 252-5210, Japan}
\affiliation{Kavli Institute for the Physics and Mathematics of the Universe (Kavli IPMU, WPI), UTIAS, The University of Tokyo, Kashiwa, Chiba 277-8583, Japan}
\affiliation{The Graduate University for Advanced Studies (SOKENDAI), Miura District, Kanagawa 240-0115, Hayama, Japan}

\author{Yuichiro Matsuzaki}
\email{ymatsuzaki872@g.chuo-u.ac.jp}
\affiliation{
Department of Electrical, Electronic, and Communication Engineering, Faculty of Science and Engineering, Chuo University, 1-13-27, Kasuga, Bunkyo-ku, Tokyo 112-8551, Japan
}

\date{\today}% It is always \today, today,

\begin{abstract}
Entanglement is a resource to improve the sensitivity of quantum sensors.
In an ideal case,
using an entangled state as a probe to detect target fields, we can beat the standard quantum limit by which all classical sensors are bounded.
However, since entanglement is fragile against decoherence, it is unclear whether entanglement-enhanced metrology is useful in a noisy environment.
Its benefit is indeed limited when estimating the amplitude of DC magnetic fields 
under the effect of parallel Markovian decoherence, where the noise operator is parallel to the target field.
In this paper, on the contrary, we show an advantage to using an entanglement over the classical strategy
under the effect of parallel Markovian decoherence when we try to detect AC magnetic fields.
We consider a scenario to induce a Rabi oscillation of the qubits with the target AC magnetic fields. Although we can, in principle, estimate the amplitude of the AC magnetic fields
from the Rabi oscillation, the signal becomes weak if the qubit frequency is significantly detuned from the frequency of the AC magnetic field.
We show that, by using the GHZ states, we can significantly enhance the signal of the detuned Rabi oscillation even under the effect of parallel Markovian decoherence.
Our method is based on the fact that the interaction time between the GHZ states and AC magnetic fields scales as $1/L$ to mitigate the decoherence effect
where $L$ is the number of qubits, which contributes to improving the bandwidth of the detectable frequencies of the AC magnetic fields.
Our results open up the way for new applications of entanglement-enhanced AC magnetometry.

\end{abstract}

\maketitle

%\tableofcontents

\section{Introduction}

Recent developments in quantum technology open up new avenues for field sensing with quantum sensors such as qubits\,\cite{doi:10.1126/science.1104149,degen2017quantum}.
These advancements have significantly enhanced the precision and capabilities of quantum sensors, making them a promising tool for various applications.
As multi-qubit systems become more accessible and sophisticated, it is interesting to explore how the uncertainty of the sensing scales with the number of qubits, $L$. 
Understanding this scaling behavior is crucial for optimizing the performance of quantum sensors and making the best use of their full potential in practical scenarios.

Let us suppose to measure DC magnetic fields. The frequencies of qubits are shifted with a static magnetic field and the shift is estimated by Ramsey spectroscopy, or equivalently, by effectively converting the shift to Pauli $X$ operators of the qubits and measuring it.
If we prepare uncorrelated $L$ qubits, the uncertainty of the measured value scales as the inverse of the square root of the number of experimental data, $L^{-1/2}$, which is called the standard quantum limit\,\cite{Giovannetti_2011}.
On the other hand, using entangled states can improve the sensitivity; with the use of the Greenberger–Horne–Zeilinger (GHZ) state\,\cite{Greenberger1989}, a maximally entangled state, the uncertainty scales as $\sim L^{-1}$. This is the celebrated Heisenberg limit\,\cite{PhysRevLett.79.3865,Giovannetti_2011}.

However, the GHZ state is highly entangled and fragile against decoherence.
It is especially vulnerable against decoherence parallel to the quantization axis 
of the qubits
\,\cite{PhysRevLett.79.3865,shaji2007qubit,demkowicz2012elusive}, although they have some tolerance against decoherence transversal to the quantization axis\,\cite{brask2015improved,chaves2013noisy,GHZ_QEC1,GHZ_QEC2,PhysRevLett.112.150801,isogawa2023vector}.
For most of the multi-qubit systems
suitable for magnetic field sensors, parallel noise remains as
one of the major obstacles to improving the sensitivity\,\cite{taylor2008high,bal2012ultrasensitive,degen2017quantum,bauch2020decoherence,hayashi2020experimental}.
In the presence of independent Markovian parallel noise, {i.e.,} if the information of each qubit individually dissipates in its local environment with noise operator that is not orthogonal to the operator of the target signal, the decoherence time of the GHZ state is $L^{-1}$ times shorter than that of the single qubit\,\cite{palma1996quantum,PhysRevLett.79.3865}. This completely spoils the advantage of the GHZ state; the scaling of the uncertainty is reduced to $L^{-1/2}$, which is the same scaling as the standard quantum limit\,\cite{PhysRevLett.79.3865,shaji2007qubit,demkowicz2012elusive}. In other words, the use of the GHZ state in DC magnetic field sensing under the effect of parallel noise
has an advantage over the separable states only when quantum noises are 
limited to specific forms
such as spatially correlated noise \cite{GHZ_spatial} or non-Markovian noise \cite{GHZ_nonMarkov1,GHZ_nonMarkov2}.

On the other hand, if one measures something other than DC magnetic field, the use of the entangled states may have advantages even in the presence of independent Markovian parallel noise.
Ref.\,\cite{GHZ_kbdyH_pbdyL} shows if we measure the effect of a multi-qubit operator, better sensitivities are achieved by the use of the GHZ states. Also, Refs.\,\cite{matsuzaki2018quantum,PhysRevApplied.16.064026} show that the GHZ states have an advantage in measuring the dissipation itself. 
However, the signal that these studies aim to measure is highly specific and the application is limited.

In this study, we propose yet another way to take advantage of the entangled state for quantum sensing in more generic situations. We consider measuring an AC signal with the GHZ state.
The sensitivity of the AC signal amplitude measurement is maximized if we know the frequency of the AC signal and also if we can use the on-resonant
condition; for example
if the frequency of the AC signal is much less than the resonant frequency of the qubit, we can use techniques such as spin echo or dynamical decoupling with a time duration corresponding to the inverse of the target frequency \,\cite{taylor2008high,balasubramanian2008nanoscale,maze2008nanoscale,de2010universal,bylander2011noise}.
Similarly, if we can set the resonant frequency of the qubit to be the same as the frequency of the AC magnetic field, we can induce the Rabi oscillation\,\cite{yoshihara2014flux,appel2015nanoscale}. In these cases, the sensitivity scales in the same way as the DC magnetic field sensing, and the GHZ state does not have an advantage over the separable states.
However, the situation changes if the on-resonant condition is not satisfied, e.g., when the qubit frequency is unable to match the signal frequency or when the signal frequency is not known in advance.
In such cases, due to the finite frequency difference between the qubit frequency and the signal frequency, the uncorrelated sensors' longer coherence time may not be fully exploited. Then, the advantage of the GHZ states can be revived.

We illustrate how the GHZ state can be advantageous by discussing a scenario where qubits with fixed frequencies are used to probe an AC signal with a known oscillating frequency through Rabi oscillation.
With the finite frequency detuning, we find that the use of the GHZ state gives better sensitivity 
than that of the separable states even under the effect of the broad range of noise models; we in particular use Markovian 
parallel
noise 
and global depolarization.
Our result shows the usefulness of the GHZ state, contrary to the widely believed idea that the sensitivity of the GHZ state is no longer superior to that of uncorrelated sensors in the presence of generic noises.

The construction of the rest of the paper is as follows. We describe the model of our interest and derive the evolution of the system of qubits under decoherence effects in Section~\ref{sec:model}. We discuss the sensitivity in terms of the signal uncertainty and compare the performance of the GHZ state and uncorrelated qubits assuming the probe qubits with a fixed frequency in Section~\ref{sec:nosweep}.
The discussion and conclusions are given in Section~\ref{sec:condis}.

\section{System and Signal} \label{sec:model}
In this section, we first introduce the qubit Hamiltonian and define the interaction between the qubits and the AC field. We then write down the Lindblad equation with quantum noises.
We consider two different procedures for the signal detection with $L$ qubits: one with using qubits individually and the other with the GHZ state. In order to compare the accuracy of the signal detection,
we introduce observables used to measure the amplitude of the signal. We derive the expectation values of the observables by solving the Lindblad equation for both individual qubits and the GHZ state with the quantum noises.

We consider a scenario where qubit sensors 
are used to detect an AC signal that uniformly affects the sensors through the Pauli $X$ operator $\sigma_X$.
We assume that the frequency and phase of the signal are known, but the amplitude
$\epsilon$ is unknown and needs to be estimated.
Also, we assume that the qubits are homogeneous and the resonant frequencies are the same. We adopt the projection measurement appropriately to estimate $\epsilon$, which we will introduce later.

The Hamiltonian of the $i$-th qubit is given by the sum of the free Hamiltonian $H_0^i$ and the interaction term of the qubit with the AC field $\Delta H^i$:
\begin{align}
    H^i&=H_0^i+\Delta H^i, \label{eq:H1}
\end{align}
where
\begin{align}
    H_0^i&=-\frac{1}{2} \omega \sigma_Z^i, \\
    \Delta H^i&= -2 \epsilon \sigma_X^i \cos m t.
\end{align}
Here, $\omega$ and $m$ are the frequencies of the qubit and the signal, respectively, and $\sigma_{X,Y,Z}^i$ are the Pauli matrices for the $i$-th qubit. 
We neglect the phase in the AC-field oscillation for simplicity.
We introduce the states of each qubit $\ket{0}$ and $\ket{1}$ as the ground and excited states of the free Hamiltonian $H_0^i$, respectively, where $\ket{0}$ and $\ket{1}$ are defined to be time-independent.
We also define $\left|\pm\right>\equiv (\left|0\right>\pm\left|1\right>)/\sqrt 2$ for later convenience.
The interaction Hamiltonian in the interaction picture for the system can be written as
\begin{align}
    H_I^i\equiv& \, e^{iH_0^i t} \Delta H^i e^{-iH_0^i t} \nonumber \\
    =&-2 \epsilon \sigma_X^i \cos\omega t \cos mt -2 \epsilon \sigma_Y^i \sin\omega t \cos mt \nonumber \\
    =&-\epsilon \sigma_X^i \left[\cos(\omega+m) t+\cos(\omega-m)t\right] \nonumber \\
    & \quad \quad - \epsilon \sigma_Y^i \left[ \sin (\omega+m)t + \sin (\omega-m)t\right]. \label{eq:HI}
\end{align}
In the following discussion, we consider the case where the signal frequency $m$ is detuned from the qubit frequency $\omega$: $\qty|m - \omega| / \omega \gtrsim 1$. We therefore do not apply the rotating wave approximation here and retain all oscillating terms in the Hamiltonian.
The Hamiltonian of the system is the sum of the Hamiltonian for each qubit:
\begin{align}
    H         =H_{0}+\Delta H,
\end{align}
where
\begin{align}
    H_{0}     &= \sum_{j=1}^L H_0^j,  \\
    \Delta H  &= \sum_{j=1}^L \Delta H^j,
\end{align}
with $L$ being the number of the qubits. The interaction Hamiltonian in the interaction picture is
\begin{align}
    H_{I}\equiv & \, e^{iH_0 t}\Delta H e^{-iH_0 t}
    =  \sum_{j=1}^L H_I^j.
\end{align}
Here, we assume that all qubits have the same oscillation frequency $\omega$ and that the effects of the external AC field are uniform across all qubits. We also assume that the interactions between qubits are negligible. 

To include the effects of various Markovian decoherence noises, we use the Lindblad equation; in the Schroedinger picture, it is given by
\begin{equation}
    \frac{d\rho (t)}{dt}= -i[H,\rho]+ D[\rho], \label{eq:linblad}
\end{equation}
where $\rho$ is the density matrix of the state and $D[\rho]$ denotes a noise superoperator.
Let us move to the interaction picture, where the density matrix operator $\rho_{I}$
is defined as
\begin{equation}
    \rho \equiv e^{-iH_0 t}\rho_I e^{iH_0 t}.
\end{equation}
The Lindblad equation becomes
\begin{align}
        \frac{d\rho_I (t)}{dt} &= -i[ H_I,\rho_I]+ D_I[\rho_I],  \label{eq:linbladint}
\end{align}
with $D_I[\rho_I]$ denoting noise superoperators in the interaction picture. 
In this paper, to demonstrate the generality of our conclusions, we consider two types of decoherence noises in the interaction picture: parallel noise and depolarizing noise.
It is worth mentioning that such parallel noise could occur for AC magnetometry
when the amplitude of the AC magnetic fields fluctuates
\cite{cai2012robust,okaniwa2024frequency}.
The noise superoperators of 
the parallel
noise and depolarizing noise are given by
\begin{align}
    D_{I,X}[\rho] &= \frac{\Gamma_X}{2} \sum_{j=1}^L \qty(\sigma_{X}^j\rho \sigma_{X}^j-\frac{1}{2} \rho), \label{eq:Dflip}\\
    D_{\text{DP}}[\rho] &= - L\Gamma_\text{DP}\rho + \frac{L\Gamma_\text{DP}}{2^L} \mathbf{1},\label{eq:DP}
\end{align}
respectively, where $\Gamma_X$ and $\Gamma_\text{DP}$ are the corresponding damping rates, and $\mathbf{1}$ is the identity operator. Since it is known that the GHZ state is robust against the transversal noise\,\cite{brask2015improved,chaves2013noisy,GHZ_QEC1,GHZ_QEC2,PhysRevLett.112.150801,isogawa2023vector},
we do not consider the noise with a Lindblad operator proportional to $\sigma_Z$ in this paper, although it may arise from, e.g., the fluctuation of the qubit frequency\,\cite{PhysRevResearch.2.033216,PhysRevA.102.043707}.
Note that the depolarization channel can be also written in terms of Lindblad operators, as discussed in the App.\,\ref{app:dp_lindblad}.

We now describe our protocol for estimating the amplitude of the signal using both uncorrelated qubit sensors and entangled qubit sensors.
In the rest of the paper, our discussion is always in the interaction
picture, and we drop the subscript $I$ of states for convenience; e.g., $\ket{+}$ at time $t$ denotes
\begin{align*}
    |{+}(t)\rangle=\frac{e^{i\omega t/2}\ket{0}+e^{-i\omega t/2}\ket{1}}{\sqrt{2}}.
\end{align*}

For observation using non-entangled individual qubits, we first prepare all qubits in the same initial state:
\begin{equation}
\ket{\psi_{\rm indv}(0)}\equiv\ket{0}=\frac{\ket{+}+\ket{-}}{\sqrt{2}}. \label{eq:0state}
\end{equation}
Second, we let the qubits evolve for a time $t$ according to the Lindblad equation, Eq.\,\eqref{eq:linbladint}.
Finally, we perform the projection measurement with the projection operator
\begin{equation}
    P_{Y}=|Y \rangle\langle Y|, \label{eq:Pindi}
\end{equation}
where
\begin{equation}
|Y \rangle\equiv\frac{1}{\sqrt{2}}\left( \ket{+}+i \ket{-}\right).
\end{equation}
The observable is the probability of the projection, which is given by
\begin{equation}
    p_{Y}\equiv \mathrm{tr} [\rho P_Y]. \label{eq:prob_indi}
\end{equation}
For the observation using the GHZ state, we first prepare the state of the qubits $\ket{\psi_{\rm GHZ}}$ in
\begin{equation}
    \ket{\psi_{\rm GHZ}(0)}=\frac{\ket{+}^{\otimes L}+ \ket{-}^{\otimes L}}{\sqrt{2}}
\end{equation}
as the initial state.
Second, we let the qubits evolve for a time $t$ according to the Lindblad equation, Eq.\,\eqref{eq:linbladint}.
Finally, we perform the projection measurement with the projection operator
\begin{equation}
    P_{{\rm GHZ},Y}=| {\rm GHZ}_Y \rangle \langle {\rm GHZ}_Y| \label{eq:PGHZ}
\end{equation}
with
\begin{equation}
    |{\rm GHZ}_Y\rangle\equiv \frac{\ket{+}^{\otimes L} +i \ket{-}^{\otimes L}}{\sqrt{2}}.
\end{equation}
Again, the observable is the probability of the projection and is given by
\begin{equation}
    p_{{\rm GHZ},Y}\equiv \mathrm{tr} [\rho P_{{\rm GHZ},Y}]. \label{eq:prob_GHZ}
\end{equation}
Note that, in these setups, setting $L=1$ in the GHZ case reproduces the result of one individual qubit in the case of uncorrelated qubits.

\subsection{Qubit system with parallel
noise}
Let us solve the Lindblad equation for the qubit system with the parallel noise. The interaction Hamiltonian, Eq.\,\eqref{eq:HI}, as well as the parallel noise, Eq.\,\eqref{eq:Dflip}, act on each qubit independently.
In such a case, i.e., when both the interaction Hamiltonian and the noise are uncorrelated, single-body operations, we can solve the Lindblad equation using the following idea. First, let us rewrite the Lindblad equation in the following form
\begin{equation}
    \frac{d \rho (t)}{dt}= \sum_{j=1}^L \qty(-i [ H_I^j,\rho]+D_j[\rho]), \label{eq:indEOM}
\end{equation}
where $D_{j}$ is the noise superoperator acting on the $j$-th qubit; e.g., for the parallel noise above,
\begin{align*}
    D_{j}[\rho] = \frac{\Gamma_X}{2} \qty(\sigma_{X}^j\rho \sigma_{X}^j-\frac{1}{2} \rho).
\end{align*}
$D_j$ acts linearly and non-trivially only on the $j$th-qubit subspace. 
Then, we decompose the density matrix into each qubit subspace.
In general, the density matrix can be written in the form of the linear sum of the tensor product of operators acting onto the subspace of individual qubits:
\begin{equation}
    \rho(t)=\sum_{n} O^{(n)}_1(t) \otimes O^{(n)}_2 (t)\otimes  \cdots \otimes O^{(n)}_L (t). \label{eq:generalexpan}
\end{equation}
Substituting this into the Lindblad equation, we obtain
\begin{equation}
\label{eq:lindblad_ind}
    \sum_{n,j} O^{(n)}_1\otimes \cdots \otimes \frac{d O^{(n)}_j}{dt} \otimes  \cdots \otimes O_L^{(n)}
    =\sum_{n,j}  O^{(n)}_1\otimes\cdots \otimes (-i[H_j,O^{(n)}_j] + D_j[O^{(n)}_j] )\otimes  \cdots\otimes O_L^{(n)}.
\end{equation}
Eq.\,\eqref{eq:lindblad_ind} holds if $O_j^{(n)}$ obeys the following equation:
\begin{equation}
    \frac{d O^{(n)}_j(t)}{ dt}= -i[H_j,O^{(n)}_j] + D_j[O^{(n)}_j]. \label{eq:EOMO}
\end{equation}
Therefore, if we decompose the initial density matrix operator into each qubit subspace and solve Eq.\,\eqref{eq:EOMO} independently, we obtain the solution of the original Lindblad equation, Eq.\,\eqref{eq:indEOM}.

Let us solve the Lindblad equation with the parallel noise using the above idea. First, let us expand the element of the density matrix of a qubit space in the $\{\ket{+},\ket{-}\}$ basis as
\begin{equation}
    \label{eq:sepbasis}
    O(t)=c_{++}(t)\ket{+}\bra{+}+c_{+-}(t)\ket{+}\bra{-}+c_{-+}(t)\ket{-}\bra{+}+c_{--}(t)\ket{-}\bra{-}.
\end{equation}
Here, $O$ itself is not a density matrix in general; its trace is not necessarily unity, nor is it required to be Hermitian. The solution of the Lindblad equation for this basis is presented in App.\,\ref{app:parallel}.

In the case of one qubit initialized as the ground state, $\ket{\psi_{\rm indv}(0)}=(\ket{+}+\ket{-})/\sqrt{2}$, we have its density matrix at a later time $t$ as
\begin{align}
    \rho_{\rm indv}(t) =& \left(\frac{1}{2}- \epsilon \int_0^t dt' \ 2\cos mt' \sin\omega t' e^{-\Gamma_X t'}\right)\ket{+}\bra{+} \nonumber \\
    & +e^{-\Gamma_X t}\left( \frac{1}{2} +i\epsilon \int_0^t dt' \ 2\cos mt' \cos \omega t' \right)\ket{+}\bra{-} \nonumber\\
    & +e^{-\Gamma_X t}\left( \frac{1}{2} - i\epsilon \int_0^t dt'\ 2\cos mt' \cos\omega t' \right)\ket{-}\bra{+} \nonumber\\
    & +\left(\frac{1}{2}+ \epsilon  \int_0^t dt' \ 2\cos mt'\sin\omega t' e^{-\Gamma_X t'} \right)\ket{-}\bra{-} .
\end{align}
The projection probability, Eq.~\eqref{eq:prob_indi}, is then given by
\begin{equation}
    p_Y=\frac{1}{2}-\epsilon t e^{-\Gamma_X t} W(t),\label{eq:indvbf}
\end{equation}
where
\begin{align}
    W(t) &\equiv \frac{1}{t} \int_0^t dt' \ 2 \cos mt' \cos\omega t' \nonumber \\
    &=\frac{ \sin \qty[(\omega+m) t ]}{(\omega+m)t} + \frac{\sin \qty[(\omega-m)t]}{(\omega-m)t}. \label{eq:window}
\end{align}
The function $W(t)$
is a window function governing the effective signal strength interacting with the sensor. 
The window function $W$ is approximately given by 
\begin{equation}
    W(t)\simeq 
    \begin{dcases}
        2  & \text{for } \ t\lesssim |\omega+m|^{-1}\\
        1 & \text{for } \ |\omega+m|^{-1} \lesssim t \lesssim |\omega-m|^{-1}\\
        \frac{1}{|\omega - m|t} & \text{for }\ |\omega-m|^{-1}\lesssim t
    \end{dcases}.
\end{equation}
Note that the rotating wave approximation corresponds to the second case above, where the evolution time is much longer than $|\omega + m|^{-1}$, but much shorter than $|\omega - m|^{-1}$, as $\omega$ is close enough to $m$.

In the case where we use the GHZ state,
the initial condition for the density matrix $\rho_{\rm GHZ}$ is given by
\begin{equation}
    \rho_{\rm GHZ}(0)=\frac{1}{2}\ket{+}^{\otimes L}\bra{+}^{\otimes L}+ \frac{1}{2}\ket{+}^{\otimes L}\bra{-}^{\otimes L}+\frac{1}{2}\ket{-}^{\otimes L}\bra{+}^{\otimes L}+\frac{1}{2}\ket{-}^{\otimes L}\bra{-}^{\otimes L}.
\end{equation}
The density matrix at a later time $t$ is then given by, according to Eq.\,\eqref{eq:evGHZbitflip},
\begin{align}
    \rho_{\rm GHZ}(t)&= \frac{1}{2}\ket{+}^{\otimes L}\bra{+}^{\otimes L}+\frac{1}{2}\ket{-}^{\otimes L}\bra{-}^{\otimes L} \nonumber \\
    &\quad + \frac{1}{2}e^{-L \Gamma_X t}\left(1+2iL\epsilon t W(t) \right)\ket{+}^{\otimes L}\bra{-}^{\otimes L} \nonumber \\
    &\quad +\frac{1}{2}e^{-L \Gamma_X t}\left(1-2iL\epsilon t W(t) \right)\ket{-}^{\otimes L}\bra{+}^{\otimes L}
    + \mathcal{O}(\epsilon^2) + \mathcal{O}(\epsilon L^0).
\end{align}
The probability of the projection, Eq.~\eqref{eq:prob_GHZ} is
\begin{equation}
    p_{{\rm GHZ},Y} \simeq \frac{1}{2}- e^{-L \Gamma_X t}L \epsilon t W(t) \label{eq:pGHZbitflip}
\end{equation}
up to the leading order in $\epsilon$ and $L$.

\subsection{Qubit system with depolarizing noise}
Next, let us solve the Lindblad equation for the qubit system with depolarizing noise, Eq.\,\eqref{eq:DP}.
Since the depolarizing noise includes the parallel noise as well, it is believed that we cannot beat the classical limit by using the GHZ state in scaling under the effect of depolarizing noise when we try to detect DC magnetic fields 
\cite{brask2015improved,demkowicz2012elusive}.
The Lindblad equation to solve is
\begin{align}
    \frac{d\rho(t)}{dt} =  -i[H_I, \rho] - L\Gamma_\text{DP}\rho + \frac{L\Gamma_\text{DP}}{2^L} \mathbf{1},
\end{align}
where $H_I$ is the interaction Hamiltonian given by Eq.~\eqref{eq:HI}.
To simplify the calculation, we focus on the dominant interaction term in $H_I$ relevant to the observable defined in Eq.~\eqref{eq:Pindi} and Eq.~\eqref{eq:PGHZ}. Specifically, we retain only the $\sigma_X$ contribution in $H_I$:
\begin{align}
    \frac{d \rho(t)}{dt} \simeq 2 i\epsilon \sum_{j=1}^L[  \sigma_X^j \cos \omega t \cos mt, \rho] - L\Gamma_\text{DP}\rho + \frac{L\Gamma_\text{DP}}{2^L} \mathbf{1}.
\end{align}

To solve the Lindblad equation, we expand the density matrix in the basis of the following $5$ operators, $\{\rho_i\} \equiv \{\rho_+, \rho_-, \rho_1, \rho_{+-},\rho_{-+}\}$;
\begin{align}
    \rho_\pm &= \tens{\ket{\pm}}{L}\tens{\bra{\pm}}{L} \\
    \rho_1 &= \frac{1}{2^{L}} (\mathbf{1} - \rho_+ - \rho_-) \\
    \rho_{\pm\mp} &= \tens{\ket{\pm}}{L}\tens{\bra{\mp}}{L}.
\end{align}
Here, $\rho_i$ are orthogonal in the sense that $\text{Tr}(\rho_i^\dagger\rho_j) \propto \delta_{ij}$.
For the GHZ state,
the initial condition of $\rho$ is
\begin{align}
    \rho_{\rm GHZ}(0) = \frac{1}{2}(\rho_+ + \rho_- + \rho_{+-} + \rho_{-+}).
    \label{rhoGHZ(0)}
\end{align}
Assuming the following ansatz for $\rho_{\rm GHZ}(t)$,
\begin{align}
    \label{eq:depoansatz}
    \rho_{\rm GHZ}(t) = \sum_i c_i(t) \rho_i,
\end{align}
we can solve the Lindblad equation as shown in App.\,\ref{app:depo}.

Then, with the measurement regarding the projection operator given in Eq.~\eqref{eq:PGHZ}, we obtain the projection probability as
\begin{align}
    p_{{\rm GHZ},Y}
    &\simeq \frac{1}{2}\left[c_+(t) + c_-(t) +i c_{+-}(t) -i c_{-+}(t)\right] \nonumber\\
    &= \frac{1}{2^L} + \left(\frac{1}{2} - \frac{1}{2^L}\right) e^{-L \Gamma_\text{DP} t} - \frac{e^{-L\Gamma_\text{DP} t}}{2}\sin (2L\epsilon t W(t)). \label{eq:pDPGHZL}
\end{align}
The above discussion holds even for the $L=1$ case. Thus, for the case of uncorrelated $L$ qubits, the projection probability for each qubit is obtained by substituting $L=1$ in Eq.\,\eqref{eq:pDPGHZL}, as
\begin{equation}
    p_{Y}\simeq\frac{1}{2}-e^{-\Gamma_\text{DP} t} \epsilon t W(t), \label{eq:indvdepo}
\end{equation}
assuming $|\epsilon t| \ll 1$.
For the GHZ state with $L \gg 1$, we obtain
\begin{equation}
    p_{{\rm GHZ},Y} \simeq \frac{1}{2}e^{-L\Gamma_\text{DP} t} -e^{-L \Gamma_\text{DP} t} L \epsilon tW(t), \label{eq:pGHZdepo}
\end{equation}
where, again, we assume $|L \epsilon t| \ll 1$.

\section{Uncertainty comparison with probe qubits with the fixed frequency
} \label{sec:nosweep}
We are now ready to compare the performance of the uncorrelated qubits and the GHZ state in the estimation of the signal amplitude $\epsilon$.
In this section, we calculate the uncertainty of the estimation of the signal amplitude, $\delta \epsilon$, and we show that under the condition that the detuning $|\omega-m|$ is large compared to the decoherence effect (of individual qubits), the performance of qubits in the GHZ state is better than uncorrelated qubits.

With the actual number of measurement data $N_{\rm r}$,
the uncertainty of the projection probability $p\, (=\text{$p_Y$ or $p_{{\rm GHZ},Y}$})$ to measure the projection operator $P$ is given by
\begin{equation}
    \delta p \equiv \sqrt{\frac{\left\langle P^2\right\rangle-\left\langle P\right\rangle^{2}}{N_\text{r}}} = \sqrt{\frac{p - p^2}{N_\text{r}}}.
\end{equation}
One can relate $\delta p$ to the uncertainty $\delta \epsilon $ of the signal amplitude $\epsilon$ as
\begin{equation}
    \delta \epsilon = \frac{\delta p}{|dp/d\epsilon|}.
    \label{eq:deltaep}
\end{equation}
To compare the performance of uncorrelated qubits and the GHZ state, we assume a fixed total observation time $T$ for both measurement schemes. 
We also fix the number of qubits $L$; for the scheme with the uncorrelated qubits, we assume $L$ qubits are
simultaneously
used,
while, for the scheme with the GHZ state,
we use the GHZ state with $L$ entangled qubits.
Therefore, the number of data $N_{\rm r}$ is given by $N_{\rm r}=L T/t$ for uncorrelated
qubits and $N_{\rm r}=T/t$ for the GHZ state, where $t$, the evolution time, is chosen independently for each measurement scheme.
We then optimize the evolution time $t$ to minimize the uncertainty $\delta \epsilon$ in the presence of parallel and depolarizing noise, 
and compare the results for uncorrelated qubits and the GHZ state.
We will see that, both for the parallel and depolarizing noise, the scheme with the GHZ state can give better sensitivity; this conclusion is drawn by semi-analytic analysis as well as numerical one.

\subsection{Parallel noise}
\subsubsection{Sensors in the separable state}
Based on the projection probability $p_Y$ given by Eq.~\eqref{eq:indvbf}, the uncertainty of the signal amplitude estimation is given by
\begin{equation}
    \delta \epsilon^{(\rm indv)}=\sqrt{\frac{t}{L T}} \frac{1/2}{e^{-\Gamma_X t } t} \frac{1}{\qty|W(t)|}. \label{eq:indvbfde}
\end{equation}

As for the dependence on the parameter $t$, we have 
for $\Gamma_X \lesssim |\omega-m|$
\begin{equation}
    \left. \delta \epsilon^{(\rm indv)}\right|_{\Gamma_X \lesssim |\omega-m|} \propto
    \begin{cases}
        {1}/{\sqrt{t}} & \text{for } t \lesssim |\omega-m|^{-1} \\
        \sqrt{t}e^{\Gamma_X t} & \text{for } |\omega-m|^{-1} \lesssim t
    \end{cases},
\end{equation}
giving the optimized time $t\simeq 1/|\omega-m|$.
On the other hand, if $\Gamma_X \gtrsim |\omega-m|$, we have
\begin{equation}
    \left. \delta \epsilon^{(\rm indv)} \right|_{\Gamma_X\gtrsim |\omega-m|} \propto
    \begin{cases}
        {e^{\Gamma_X t}}/{\sqrt{t}} & \text{for } t \lesssim |\omega-m|^{-1}\\
        \sqrt{t}e^{\Gamma_X t} & \text{for } |\omega-m|^{-1} \lesssim t
    \end{cases},
\end{equation}
giving the optimized time $t\simeq\frac{1}{2\Gamma_{X}}$.
As a result, with the optimized time chosen, we have the following uncertainty
\begin{equation}
    \delta \epsilon^{(\rm indv)} \simeq
    \begin{dcases}
        \sqrt{\frac{|\omega-m|}{T}} \frac{1}{\sqrt{L}} & \text{for } \Gamma_X \lesssim |\omega-m|\\
        \sqrt{\frac{\Gamma_X}{T}}\frac{1}{\sqrt{L}} & \text{for } \Gamma_X \gtrsim |\omega-m|
    \end{dcases}
\end{equation}
up to $\mathcal{O}(1)$ constant factors.

\subsubsection{Sensors in the GHZ state}
Based on the projection probability $p_{{\rm GHZ},Y}$ given by Eq.~\eqref{eq:pGHZbitflip},
the uncertainty reads as
\begin{equation}
    \delta \epsilon^{\rm (GHZ)}\simeq \sqrt{\frac{t}{T}}\frac{1/2}{e^{-L\Gamma_X t } tL} \frac{1}{\qty|W(t)|}. \label{eq:GHZbfde}
\end{equation}

Similar to the previous analysis for the uncorrelated
qubit case, based on time $t$ dependence, we obtain that
for $L\Gamma_X \lesssim |\omega-m|$ the optimized time is $t\simeq 1/|\omega-m|$, while at a large $L$ with $L\Gamma_X \gtrsim |\omega-m|$,
the optimized evolution time is around $t\simeq0.5/L\Gamma_X$, which is, just as in the case of DC signals, $L$ times shorter than the time for the uncorrelated qubits. 
As a result, with an optimized choice of time $t$, we obtain
\begin{equation}
    \label{eq:GHZbfdeApprox}
    \delta \epsilon^{ \rm (GHZ)}\simeq 
    \begin{dcases}
        \sqrt{\frac{|\omega-m|}{T}} \frac{1}{L} & \text{for } L\Gamma_X \lesssim |\omega-m| \\
        \sqrt{\frac{\Gamma_X}{T}} \frac{1}{\sqrt{L}} & \text{for } L\Gamma_X \gtrsim |\omega-m|
    \end{dcases}.
\end{equation}
up to $\mathcal{O}(1)$ constant factors.
We emphasize that, in the case of $\Gamma_X\lesssim |\omega -m|$, the GHZ state gives better sensitivity. For $L\Gamma_X\lesssim |\omega -m|$ and 
$\Gamma_X\lesssim |\omega -m|\lesssim L\Gamma_X$, the ratio $\delta\epsilon^{\rm (GHZ)}/\delta\epsilon^{\rm (indiv)}$ is 
$\mathcal{O}(1/\sqrt{L})$ and
$\mathcal{O}(\Gamma_X^{1/2}/|\omega-m|^{1/2})$, respectively. Accordingly, the sensitivity with the GHZ state can be orders of magnitude better than that with the separable state, as we will show below.
It is worth mentioning that
our results are different from the Heisenberg-limited scaling, because $\delta \epsilon^{ \rm (GHZ)}$ scales as $1/\sqrt{L}$ in the limit of large $L$.

\begin{figure}[t]
  \centering
  \includegraphics[width=0.6 \linewidth]{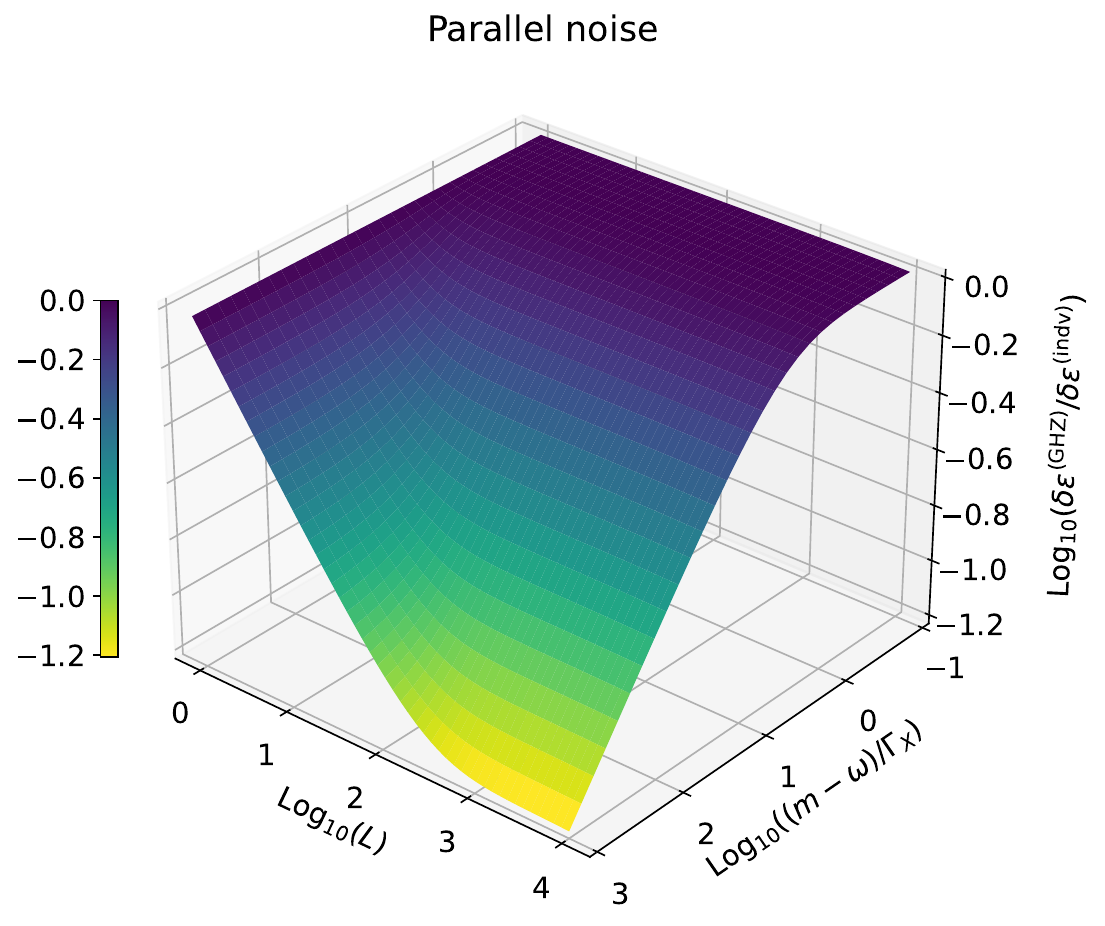}
  \caption{3D plot of the ratio $\delta \epsilon^{\rm (GHZ)}/\delta \epsilon^{\rm (indv)}$ between the GHZ case (Eq.~\eqref{eq:GHZbfde}) and uncorrelated qubits (Eq.~\eqref{eq:indvbfde}) case under the decoherence due to the parallel noise.  The numerically optimized times are chosen for each case for given parameters $(L,m)$. The decoherence rate is fixed as $\Gamma_X/\omega=10^{-6}$. The color bar shows the value of $\log(\delta \epsilon^{\rm (GHZ)}/\delta \epsilon^{\rm (indv)})$.}
  \label{fig:3Dplotbf}
\end{figure}
%%
%%%
\begin{figure}[t]
  \centering
  \includegraphics[width=0.6 \linewidth]{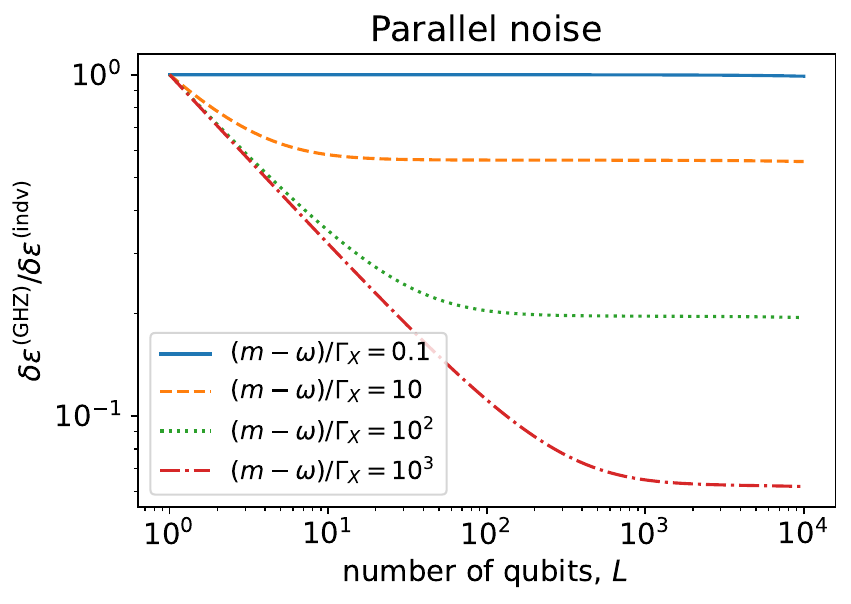}
  \caption{{The ratio $\delta \epsilon^{\rm (GHZ)}/\delta \epsilon^{\rm (indv)}$ as a function of number of qubits $L$ using Eqs.~\eqref{eq:GHZbfde} and \eqref{eq:indvbfde}. The decoherence due to the parallel noise is applied and the numerically-optimized times are chosen for the given values of signal frequency $m$. The decoherence rate is fixed as $\Gamma_X/\omega=10^{-6}$.}}
  \label{fig:lineplotbf}
\end{figure}

In Figs.~\ref{fig:3Dplotbf} and \ref{fig:lineplotbf}, we plot
the ratio $\delta \epsilon^{\rm (GHZ)}/\delta \epsilon^{\rm (indv)}$ as a function
of $L$ and the detuning $m-\omega$ (which is taken to be positive), and as a function of $L$ with several given values of $m-\omega$, respectively.
Here, we fix $\omega$ and take $\Gamma_X/\omega=10^{-6}$. Such a choice of $\Gamma_X$ is just for the presentation. Our conclusions are unchanged irrespective of the value of $\Gamma_X$ and hold for a larger or smaller value of $\Gamma_X$.
From the figures, we can see that the performance of the GHZ state is limited when the detuning $|\omega-m|$ is not much larger than the decoherence rate $\Gamma_{X}$, as previously known. However, as the detuning becomes larger, the GHZ state outperforms the uncorrelated qubits with an increasing number of qubits $L$ because the optimized evolution time is determined by the detuning. With large $L$ such that $L\Gamma_X \gtrsim |\omega - m|$, the decoherence effect becomes dominant again for the GHZ state, and the performance of the GHZ state is saturated.

\subsection{Depolarizing noise}
\subsubsection{Sensors in the separable state}
With the number of data $N_{\rm r}=L T/t$ and the observable $p_Y$ given by Eq.~\eqref{eq:indvdepo}, the uncertainty is the same as the case of parallel noise with the replacement $\Gamma_X\rightarrow \Gamma_\text{DP}$:
\begin{equation}
    \delta \epsilon^{(\rm indv)}=\sqrt{\frac{t}{L T}} \frac{1/2}{e^{-\Gamma_\text{DP} t } t} \frac{1}{\qty|W(t)|}. \label{eq:indvdepode}
\end{equation}For $\Gamma_\text{DP}\lesssim|\omega-m|$ the optimized time is $t\simeq |\omega-m|^{-1}$ while for $\Gamma_\text{DP}\gtrsim|\omega-m|$ the optimized time is $t\simeq\frac{1}{2\Gamma_\text{DP}}$. Therefore, we obtain
\begin{equation}
    \delta \epsilon^{(\rm indv)} \simeq
    \begin{dcases}
        \sqrt{\frac{|\omega-m|}{T}} \frac{1}{\sqrt{L}} & \text{for } \Gamma_\text{DP} \lesssim |\omega-m|\\
        \sqrt{\frac{\Gamma_\text{DP}}{T}}\frac{1}{\sqrt{L}} & \text{for } \Gamma_\text{DP} \gtrsim |\omega-m|
    \end{dcases},
\end{equation}
up to $\mathcal{O}(1)$ constant factors.

\subsubsection{Sensors in the GHZ state}
With the repetition time $N_{\rm r}=T/t$ and 
the projection probability given by Eq.~\eqref{eq:pGHZdepo}, we obtain the uncertainty of the signal amplitude as
\begin{equation}
    \delta \epsilon^{\rm(GHZ)}\simeq\sqrt{\frac{t}{T}}\frac{\sqrt{e^{L\Gamma_\text{DP} t}/2-1/4}}{ tL} \frac{1}{\qty|W(t)|}. \label{eq:GHZdepode}
\end{equation}
Similarly to the discussion of 
parallel noise, we can find that the optimized time is $t\simeq |\omega-m|^{-1}$ for $L\Gamma_\text{DP} \lesssim |\omega-m|$, while for $L\Gamma_\text{DP} \gtrsim |\omega-m|$, $\delta \epsilon^{\rm(GHZ)}$ is minimized when $t\simeq \frac{0.8}{\Gamma_\text{DP} L}$. Therefore, the uncertainty at the optimized choice of time $t$ reads as
\begin{equation}
    \label{eq:GHZdepodeApprox}
    \delta \epsilon^{ \rm (GHZ)}\simeq 
    \begin{dcases}
        \sqrt{\frac{|\omega-m|}{T}} \frac{1}{L} & \text{for } L\Gamma_\text{DP} \lesssim |\omega-m| \\
        \sqrt{\frac{\Gamma_\text{DP}}{T}} \frac{1}{\sqrt{L}} & \text{for } L\Gamma_\text{DP} \gtrsim |\omega-m|
    \end{dcases},
\end{equation}
up to $\mathcal{O}(1)$ constant factors.
For $L\Gamma_\text{DP} \lesssim |\omega-m|$, the uncertainty also scales as $\delta\epsilon^{(\rm GHZ)} \propto L^{-1}$.

%%%
\begin{figure}[t]
  \centering
  \includegraphics[width=0.6 \linewidth]{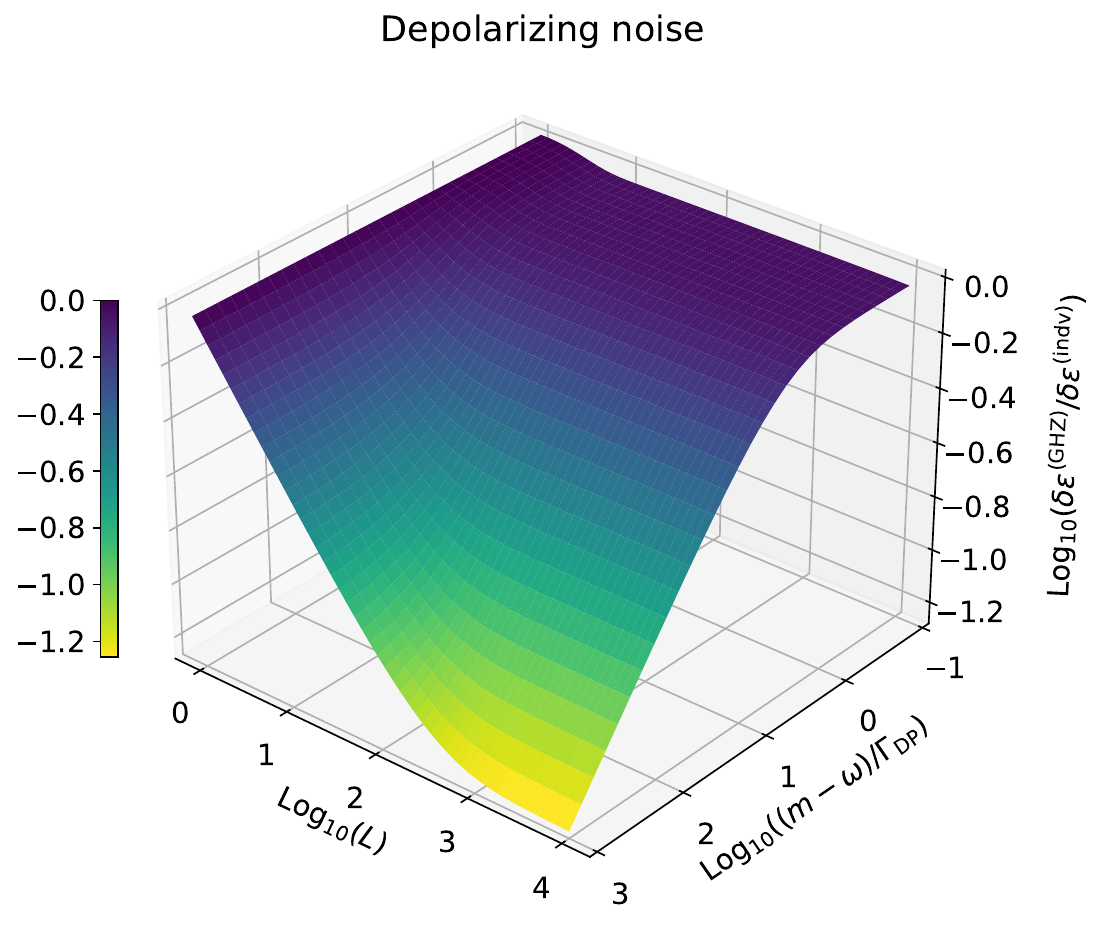}
  \caption{3D plot of the ratio $\delta \epsilon^{\rm (GHZ)}/\delta \epsilon^{\rm (indv)}$ between the GHZ case (Eq.~\eqref{eq:GHZdepode}) and uncorrelated qubits (Eq.~\eqref{eq:indvdepode}) case under depolarizing decoherence and with numerically-optimized times chosen for given parameters $(L,m)$. The decoherence rate is fixed as $\Gamma_{\rm DP}/\omega=10^{-6}$.  The color bar shows the value of $\log(\delta \epsilon^{\rm (GHZ)}/\delta \epsilon^{\rm (indv)})$.}
  \label{fig:3Dplotdepo}
\end{figure}
%%
%%%
\begin{figure}[t]
  \centering
  \includegraphics[width=0.6 \linewidth]{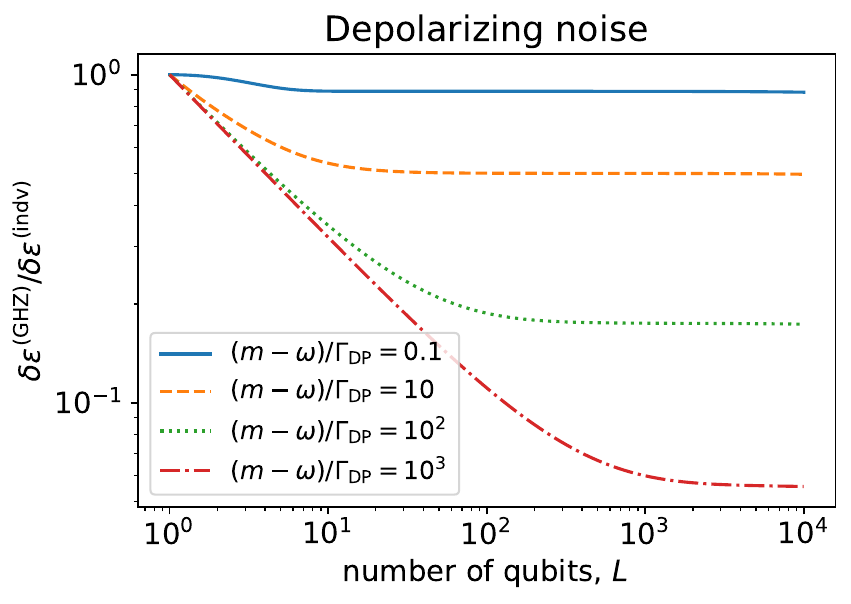}
  \caption{{The ratio $\delta \epsilon^{\rm (GHZ)}/\delta \epsilon^{\rm (indv)}$ as a function of number of qubits $L$ using Eqs.~\eqref{eq:GHZdepode} and \eqref{eq:indvdepode}. Depolarizing decoherence is applied and the optimized times are chosen for both cases for a given value of signal frequency $m$. The decoherence rate is fixed as $\Gamma_{\rm DP}/\omega=10^{-6}$.}}
  \label{fig:lineplotdepo}
\end{figure}

In Fig.~\ref{fig:3Dplotdepo} and Fig.~\ref{fig:lineplotdepo}, we present the ratio $\delta \epsilon^{\rm (GHZ)}/\delta \epsilon^{\rm (indv)}$ as a function of $L$ and the detuning $m-\omega$ (which is taken to be positive), and as a function of $L$ for several fixed values of $m-\omega$, respectively. In both figures, $\omega$ is fixed and $\Gamma_{\rm DP}$ is given as $\Gamma_{\rm DP}/\omega=10^{-6}$. For each case, the evolution time is numerically optimized to maximize the sensitivity.
Similar to the parallel noise scenario, the GHZ state demonstrates superior performance over uncorrelated qubits as the number of qubits $L$ increases, when the detuning $|\omega-m|$ is large enough.
On the other hand,  when the detuning is small, the performance of using the GHZ states is limited, which is consistent with the previous results discussing the use of the GHZ states to detect DC magnetic fields under the effect of 
depolarizing noise\,\cite{brask2015improved,demkowicz2012elusive}.

\subsection{Other noises}
The analytic results of the uncertainty ratios between the GHZ state and the uncorrelated qubits are qualitatively similar for both parallel
and depolarizing noise:
\begin{equation}
    \frac{\delta \epsilon^{\rm (GHZ)}}{\delta \epsilon^{\rm (indv)}}\simeq
    \begin{dcases}
        1                                & \text{for }|\omega-m|\lesssim \Gamma                    \\
        \sqrt{\frac{\Gamma}{|\omega-m|}} & \text{for }\Gamma \lesssim |\omega-m| \lesssim L \Gamma \\
        \frac{1}{\sqrt{L}}               & \text{for }L\Gamma \lesssim |\omega-m|                  \\
    \end{dcases},
    \label{eq:ratiofixedprobe}
\end{equation}
where $\Gamma$ is $\Gamma_{X}$ and $\Gamma_\text{DP}$ for the
parallel case and depolarizing case, respectively. The numerical results shown in Figs.~\ref{fig:3Dplotbf}, \ref{fig:lineplotbf}, \ref{fig:3Dplotdepo}, and \ref{fig:lineplotdepo} are consistent with the scaling and saturation behavior with respect to the number of qubits $L$ and the signal frequency $m$.
This comes from the fact that both models show an exponential decay of the quantum states.

From these results, we infer that this scaling behavior is likely applicable to a broad range of noise models which show an exponential decay. That is, when the detuning between the probing qubit frequency $\omega$ and the signal frequency $m$ is smaller than the decoherence rate of individual qubits, the GHZ state does not offer any advantage over uncorrelated qubits against parallel noise, as previously established\,\cite{PhysRevLett.79.3865,shaji2007qubit,demkowicz2012elusive}. However, as the detuning increases, the optimal evolution time is dictated by the detuning rather than the decoherence rate. In the regime of large detuning $|\omega-m| \gtrsim L \Gamma$, where $\Gamma$ represents the decoherence rate, the impact of decoherence becomes negligible compared to the detuning. Consequently, the GHZ state surpasses uncorrelated qubits with a scaling advantage of $1/\sqrt{L}$.
In essence, with finite detunings, the ability of each qubit to retain information for extended periods is not fully exploited. Nonetheless, the GHZ state can benefit from the entanglement among qubits.

It is important to note that while the GHZ state shows superior performance over uncorrelated qubits in the presence of detuning, this does not imply that the GHZ state reaches the Heisenberg limit or surpasses the standard quantum limit.
 Even the \emph{tuned} GHZ state, i.e., the GHZ state with qubits whose frequency matches that of the signal, cannot reach the Heisenberg limit in the presence of generic noises\,\cite{PhysRevLett.79.3865,shaji2007qubit,demkowicz2012elusive} and their sensitivities are no better than that of the \emph{tuned} uncorrelated qubits.
The optimal sensitivity of the \emph{detuned} GHZ state, as indicated in Eq.\,\eqref{eq:GHZbfdeApprox} and Eq.\,\eqref{eq:GHZdepodeApprox}, remains comparable to the standard quantum limit of \emph{tuned} uncorrelated qubits.
Thus, the \emph{detuned} GHZ state outperforms \emph{detuned} uncorrelated qubits, but not their \emph{tuned} counterparts. 

\section{Discussion and Conclusion} \label{sec:condis}

In this paper, we have explored the performance of quantum sensors in detecting an AC oscillating signal. Assuming the signal's frequency and phase are known, our objective was to estimate its amplitude. We compared the sensitivity using GHZ states with
that using uncorrelated qubits under the influence of 
Markovian parallel
and depolarizing noise. With fixed qubit frequencies and potential detuning from the signal frequency, we calculated and compared the uncertainty in amplitude estimation for both GHZ states and uncorrelated qubits.

Our main finding is that the GHZ state surpasses uncorrelated qubits in estimating the signal amplitude when the detuning between the qubit frequency and the signal frequency is significantly larger than the decoherence rate of individual qubits. In such scenarios, the optimal evolution time is dictated by the detuning rather than the decoherence rate, rendering the decoherence effect negligible. This allows us to make use of the entanglement among qubits to enhance sensitivity. Conversely, when the detuning is small relative to the decoherence rate, the GHZ state's performance is constrained by
decoherence due to Markovian parallel and depolarizing noises, offering no advantage over uncorrelated qubits, as previously established. It is important to note that the GHZ state does not achieve the Heisenberg limit or exceed the standard quantum limit. Instead, we have demonstrated an additional benefit of the GHZ state over uncorrelated qubits when detuning is considered.

Our setup applies to various real-world scenarios. For instance, even if the signal's frequency is known, the qubit's frequency might be fixed, have a limited tunable range, or tuning might significantly degrade coherence time. In such situations, employing GHZ states can potentially enhance the sensitivity of the qubits. Additionally, if the exact frequency of the signal is unknown but a rough estimate is available, GHZ states can be used to probe a wide range of frequencies. This means GHZ states can be employed for preliminary checks to detect the presence of a signal before a more detailed search, or to survey a broad frequency range.

We explain possible physical realization.
Electron spin ensembles such as donors in silicon material \cite{tyryshkin2012electron} and nitrogen-vacancy (NV) centers in diamond \cite{wolf2015subpicotesla} are some of the candidates 
for realizing entanglement-enhanced sensors. 
Electron spins interact with the magnetic fields, and the resonant frequency shifts due to the Zeeman energy, which is useful for detecting magnetic fields \cite{degen2017quantum}.
A superconducting qubit can be collectively coupled with the electron spin ensembles \cite{zhu2011coherent,kubo2011hybrid,zhu2014observation}, potentially providing a way to generate entanglement between the electron spins \cite{tanaka2015proposed,dooley2016hybrid,tatsuta2024generation}. There is a theoretical proposal for the NV centers in diamonds to generate multi-partite entanglement using the dipole-dipole interaction by applying a global microwave \cite{matsuzaki2022generation}. Such a property is a prerequisite to realizing the entanglement-enhanced sensors.

The scenario to survey a broad frequency range is particularly relevant for the search for extremely weak signals, opening up new possibilities for the application of quantum sensors in high energy physics\,\cite{Chou:2023hcc}, such as high-sensitivity measurements for dark matter signals\,\cite{Dixit:2020ymh,Chen:2022quj,Chen:2023swh,Chen:2024aya,Chigusa:2023hms,Chigusa:2024psk,Ito:2023zhp} or gravitational waves\,\cite{Ballmer:2022uxx,Ito:2023bnu}. In these cases, the possible frequency range of the target signal is vast, and detection requires the highest sensitivity. When a large number of quantum sensors are available, using quantum sensors in the GHZ state, even with or without tunability of the qubit frequency, might be a promising method for searching for such extremely weak signals.
For light dark matter detection, in particular, the signal frequency is unknown. Hence, we need to scan a wide frequency range to detect a signal. This scanning procedure is usually time-consuming and
one of the significant obstacles to detecting a dark matter signal. Our findings indicate that using the GHZ state as a quantum sensor may greatly reduce the scan time to cover the wide frequency range, compared with the case of individual qubits.

We have focused on estimating the signal amplitude, assuming the signal frequency is known. However, it is also crucial to consider scenarios where the signal frequency is unknown, requiring us to search for both the signal amplitude and frequency using qubits with tunable frequencies.
In such cases, we can ``scan'' a finite frequency range to identify the signal frequency by performing measurements at various frequencies and adjusting the qubit frequency accordingly. Our results indicate that the GHZ state is less sensitive to detuning and can probe a wide range of frequencies simultaneously. Therefore, using the GHZ state can be advantageous for searching and measuring the signal amplitude.
However, the GHZ state's insensitivity to detuning might be a drawback when measuring the signal frequency with high precision. This suggests that while the GHZ state is more sensitive to the signal amplitude, it sacrifices sensitivity to the signal frequency, although different ``scanning'' protocols, such as a bisectional search over the frequency range, might be considered to improve sensitivity to both the signal amplitude and frequency. 
We leave this as future work.

%%%%%%%%%%%%%%%%%%%%%%%%%%%%%%%%%%%%%%%%%%%%
%\acknowledgments
%%%%%%%%%%%%%%%%%%%%%%%%%%%%%%%%%%%%%%%%%%%%
\acknowledgments
%The work of H.F. was supported by JSPS KAKENHI Grant No.\ 24K17042.
The work of T.S. was supported by the JSPS fellowship Grant No.\ 23KJ0678.
This work was supported by JSPS KAKENHI Grant Nos.\ 24K17042 [H.F.], 
23K22486 [T.M.], 	20H05661 [Y.M.], 23H04390 [Y.M.], and
24K07010 [K.N.].
This work was supported by World Premier International Research Center Initiative (WPI), MEXT, Japan.
This work was supported by
JST Moonshot (Grant Number JPMJMS226C) and CREST
(JPMJCR23I5), JST.
This work was supported by the Simons Foundation [S.C.].
This material is based upon work supported by the U.S. Department of Energy, Office of Science under grant Contract Number DE-AC02-05CH11231, partially through Quantum Information Science Enabled Discovery (QuantISED) for High Energy Physics (KA2401032), and the BNL C2QA award under grant Contract Number DE-SC0012704 (SUBK\# 390034) [S.C.].

%%%%%%%%%%%%%%%
\appendix
%%%%%%%%%%%%%%%
\section{Lindblad operators for the depolarization channel}
\label{app:dp_lindblad}
In this appendix, we derive the Lindblad operators for the depolarization channel, Eq.\,\eqref{eq:DP}. First, let us define the following operation $\mathcal{E}$;
\begin{align}
    \mathcal{E}\qty(\mathcal{A}) = \frac{1}{4^L}\sum_{\vec{v}} \sigma_{v_1}^1 \sigma_{v_2}^2 \cdots \sigma_{v_L}^L \mathcal{A} \sigma_{v_1}^1 \sigma_{v_2}^2 \cdots \sigma_{v_L}^L,
\end{align}
where $\vec{v}=(v_1,v_2,\cdots,v_L)$, $v_i \in\{0, 1, 2, 3\}$, and $\sigma_0^i=\mathbf{1}^i$, $\sigma_1^i=\sigma_X^i$, $\sigma_2^i=\sigma_Y^i$, and $\sigma_3^i=\sigma_Z^i$. The sum is taken over all possible $\vec{v}$,  i.e., $4^L$ terms.
We can now find
\begin{align}
    \mathcal{E}\qty(\mathbf{1}) &= \mathbf{1}, \\
    \mathcal{E}\qty(\sigma_{v_1}^1 \sigma_{v_2}^2 \cdots \sigma_{v_L}^L) &= 0, \quad \text{for } \vec{v} \neq 0.
\end{align}
Since we can expand any density matrix operator $\rho$ as
\begin{align}
    \rho = \frac{1}{2^L} \mathbf{1} + \sum_{\vec{v}\neq 0} \rho_{\vec{v}} \sigma_{v_1}^1 \sigma_{v_2}^2 \cdots \sigma_{v_L}^L,
\end{align}
using some $\rho_{\vec{v}}$, we can find
\begin{align}
    \frac{1}{2^L} \mathbf{1} = \mathcal{E}(\rho).
\end{align}
for any $\rho$. Therefore,
\begin{align}
    D_{\text{DP}}[\rho] &= - L\Gamma_\text{DP}\rho + \frac{L\Gamma_\text{DP}}{2^L} \mathbf{1} \nonumber\\
    &= - L\Gamma_\text{DP}\rho + L\Gamma_\text{DP} \mathcal{E}(\rho) \nonumber \\
    &= \sum_{\vec{v} \neq 0} \qty(L_{\vec{v}} \rho L_{\vec{v}}^\dagger - \frac{1}{2}\qty{L_{\vec{v}}^\dagger L_{\vec{v}}, \rho}),
\end{align}
where we have defined
\begin{align}
    L_{\vec{v}} = \frac{\sqrt{L\Gamma_\text{DP}}}{2^L} \sigma_{v_1}^1 \sigma_{v_2}^2 \cdots \sigma_{v_L}^L.
\end{align}
The dependence on $L$ in Eq.\,\eqref{eq:DP} is chosen so that the uncertainty of the signal with the GHZ state, $\delta \epsilon^{\rm (GHZ)}$, is reduced to the same order as that with the uncorrelated qubits, $\delta \epsilon^{\rm (indv)}$, when the signal frequency is exactly matched with the qubit frequency. 

\section{Solution of the Lindblad equation for the parallel noise}
\label{app:parallel}
In this appendix, we show the solution of the Lindblad equation for the parallel noise with the basis of the density matrix, Eq.\,\eqref{eq:sepbasis}.
The Lindblad equation for the element of the density matrix, Eq.\,\eqref{eq:EOMO},
reads as
\begin{align}
    \frac{dc_{++}}{dt} & = -2 \epsilon \sin \omega t \cos mt \qty(c_{+-}+ c_{-+}),                                                           \\
    \frac{dc_{--}}{dt} & = 2 \epsilon \sin \omega t \cos mt \qty(c_{+-}+ c_{-+}),                                                            \\
    \frac{dc_{+-}}{dt} & = (4 i \epsilon \cos \omega t \cos mt - \Gamma_{X}) c_{+-}+ 2 \epsilon \sin \omega t \cos mt \qty(c_{++}- c_{--}),  \\
    \frac{dc_{-+}}{dt} & = (-4 i \epsilon \cos \omega t \cos mt - \Gamma_{X}) c_{-+}+ 2 \epsilon \sin \omega t \cos mt \qty(c_{++}- c_{--}).
\end{align}
At the leading order in $\epsilon$, we obtain
\begin{subequations}
    \begin{align}
        c_{++}(t) & = c_{++}(0) - 2 \epsilon \int_{0}^{t} dt' \cos mt' \sin \omega t' e^{-\Gamma_X t'}\qty[c_{+-}(0)+c_{-+}(0)],                  \\
        c_{--}(t) & = c_{--}(0) + 2 \epsilon \int_{0}^{t} dt' \cos mt' \sin \omega t' e^{-\Gamma_X t'}\qty[c_{+-}(0)+c_{-+}(0)],                  \\
        c_{+-}(t) & = e^{-\Gamma_X t}\qty(1 + 4 i \epsilon \int_{0}^{t} dt' \cos mt' \cos \omega t') c_{+-}(0) \nonumber                          \\
                  & \quad \quad + 2 \epsilon e^{-\Gamma_X t}\int_{0}^{t} dt' \cos(mt') \sin (\omega t') e^{\Gamma_X t'}\qty[c_{++}(0)-c_{--}(0)], \\
        c_{-+}(t) & = e^{-\Gamma_X t}\qty(1 - 4 i \epsilon \int_{0}^{t} dt' \cos mt' \cos \omega t') c_{-+}(0) \nonumber                          \\
                  & \quad \quad + 2 \epsilon e^{-\Gamma_X t}\int_{0}^{t} dt' \cos(mt') \sin (\omega t') e^{\Gamma_X t'}\qty[c_{++}(0)-c_{--}(0)].
    \end{align}
\end{subequations}
For later convenience, we note the evolution of each component in $O$:
\begin{subequations}
    \label{eq:evGHZbitflip}
    \begin{align}
        \ket{+}\bra{+}\Rightarrow & \, \ket{+}\bra{+}+ 2 \epsilon e^{-\Gamma_X t}\int_{0}^{t} dt' \cos mt' \sin \omega t' e^{\Gamma_X t'}(\ket{+}\bra{-}+\ket{-}\bra{+}), \\
        \ket{-}\bra{-}\Rightarrow & \, \ket{-}\bra{-}- 2 \epsilon e^{-\Gamma_X t}\int_{0}^{t} dt' \cos mt' \sin \omega t' e^{\Gamma_X t'}(\ket{+}\bra{-}+\ket{-}\bra{+}), \\
        \ket{+}\bra{-}\Rightarrow & \, e^{-\Gamma_X t}\left( 1 +4i\epsilon \int_{0}^{t} dt' \cos mt' \cos\omega t' \right) \ket{+}\bra{-}\nonumber                        \\
                                  & \quad\quad -2\epsilon \int_{0}^{t} dt' \cos mt' \sin\omega t' e^{-\Gamma_X t'}(\ket{+}\bra{+}-\ket{-}\bra{-}),                        \\
        \ket{-}\bra{+}\Rightarrow & \, e^{-\Gamma_X t}\left( 1 -4i\epsilon \int_{0}^{t} dt' \cos mt' \cos\omega t' \right) \ket{-}\bra{+}\nonumber                        \\
                                  & \quad\quad -2\epsilon \int_{0}^{t} dt' \cos mt' \sin \omega t' e^{-\Gamma_X t'}(\ket{+}\bra{+}-\ket{-}\bra{-}).
    \end{align}
\end{subequations}

\section{Solution of the Lindblad equation for the depolarizing noise}
\label{app:depo}
Using the ansatz, Eq.\,\eqref{eq:depoansatz},
the Lindblad equation is reduced to
\begin{align}
    \sum_{i} \frac{d c_{i}}{d t}\rho_{i} = & i (2\epsilon \cos(\omega t) \cos(mt)) (2Lc_{+-}\rho_{+-}- 2Lc_{-+}\rho_{-+}) \nonumber                                             \\
                                           & - L\Gamma_{\text{DP}}\sum_{i} c_{i}(t) \rho_{i} + L\Gamma_{\text{DP}}\left[\rho_{1} + \frac{1}{2^{L}}(\rho_{+} + \rho_{-})\right].
\end{align}
Using the orthogonality of $\rho_{i}$, we obtain the following differential equations
for $c_{i}(t)$:
\begin{align}
    \frac{d c_{1}}{d t}      & = L\Gamma_{\text{DP}}(1 - c_{1}),                                                    \\
    \frac{d c_{\pm}}{d t}    & = -L\Gamma_{\text{DP}}c_{\pm} + \frac{L\Gamma_{\text{DP}}}{2^{L}},                   \\
    \frac{d c_{\pm\mp}}{d t} & = \left(\pm 4L\epsilon i \cos\omega t\cos mt - L\Gamma_{\text{DP}}\right)c_{\pm\mp}.
\end{align}
With the initial condition given in Eq.\ \eqref{rhoGHZ(0)}, the solution is given
as
\begin{align}
    c_{1}(t)   & = 1 - e^{-L\Gamma_\text{DP} t},                                                       \\
    c_{\pm}(t) & = \frac{1}{2^{L}}+ \left(\frac{1}{2}- \frac{1}{2^{L}}\right)e^{-L\Gamma_\text{DP} t}, \\
    c_{+-}(t)  & = \frac{1}{2}e^{L\left(2\epsilon W(t) i - \Gamma_\text{DP} \right)t},                 \\
    c_{-+}(t)  & = c_{+-}^{\star},
\end{align}
where the window function $W(t)$ is defined by Eq.~\eqref{eq:window}.

\bibliography{papers}% Produces the bibliography via BibTeX.

\begin{thebibliography}{53}%
\makeatletter
\providecommand \@ifxundefined [1]{%
 \@ifx{#1\undefined}
}%
\providecommand \@ifnum [1]{%
 \ifnum #1\expandafter \@firstoftwo
 \else \expandafter \@secondoftwo
 \fi
}%
\providecommand \@ifx [1]{%
 \ifx #1\expandafter \@firstoftwo
 \else \expandafter \@secondoftwo
 \fi
}%
\providecommand \natexlab [1]{#1}%
\providecommand \enquote  [1]{``#1''}%
\providecommand \bibnamefont  [1]{#1}%
\providecommand \bibfnamefont [1]{#1}%
\providecommand \citenamefont [1]{#1}%
\providecommand \href@noop [0]{\@secondoftwo}%
\providecommand \href [0]{\begingroup \@sanitize@url \@href}%
\providecommand \@href[1]{\@@startlink{#1}\@@href}%
\providecommand \@@href[1]{\endgroup#1\@@endlink}%
\providecommand \@sanitize@url [0]{\catcode `\\12\catcode `\$12\catcode `\&12\catcode `\#12\catcode `\^12\catcode `\_12\catcode `\%12\relax}%
\providecommand \@@startlink[1]{}%
\providecommand \@@endlink[0]{}%
\providecommand \url  [0]{\begingroup\@sanitize@url \@url }%
\providecommand \@url [1]{\endgroup\@href {#1}{\urlprefix }}%
\providecommand \urlprefix  [0]{URL }%
\providecommand \Eprint [0]{\href }%
\providecommand \doibase [0]{http://dx.doi.org/}%
\providecommand \selectlanguage [0]{\@gobble}%
\providecommand \bibinfo  [0]{\@secondoftwo}%
\providecommand \bibfield  [0]{\@secondoftwo}%
\providecommand \translation [1]{[#1]}%
\providecommand \BibitemOpen [0]{}%
\providecommand \bibitemStop [0]{}%
\providecommand \bibitemNoStop [0]{.\EOS\space}%
\providecommand \EOS [0]{\spacefactor3000\relax}%
\providecommand \BibitemShut  [1]{\csname bibitem#1\endcsname}%
\let\auto@bib@innerbib\@empty
%</preamble>
\bibitem [{\citenamefont {Giovannetti}\ \emph {et~al.}(2004)\citenamefont {Giovannetti}, \citenamefont {Lloyd},\ and\ \citenamefont {Maccone}}]{doi:10.1126/science.1104149}%
  \BibitemOpen
  \bibfield  {author} {\bibinfo {author} {\bibfnamefont {Vittorio}\ \bibnamefont {Giovannetti}}, \bibinfo {author} {\bibfnamefont {Seth}\ \bibnamefont {Lloyd}}, \ and\ \bibinfo {author} {\bibfnamefont {Lorenzo}\ \bibnamefont {Maccone}},\ }\bibfield  {title} {\enquote {\bibinfo {title} {Quantum-enhanced measurements: Beating the standard quantum limit},}\ }\href {\doibase 10.1126/science.1104149} {\bibfield  {journal} {\bibinfo  {journal} {Science}\ }\textbf {\bibinfo {volume} {306}},\ \bibinfo {pages} {1330--1336} (\bibinfo {year} {2004})},\ \Eprint {http://arxiv.org/abs/https://www.science.org/doi/pdf/10.1126/science.1104149} {https://www.science.org/doi/pdf/10.1126/science.1104149} \BibitemShut {NoStop}%
\bibitem [{\citenamefont {Degen}\ \emph {et~al.}(2017)\citenamefont {Degen}, \citenamefont {Reinhard},\ and\ \citenamefont {Cappellaro}}]{degen2017quantum}%
  \BibitemOpen
  \bibfield  {author} {\bibinfo {author} {\bibfnamefont {Christian~L}\ \bibnamefont {Degen}}, \bibinfo {author} {\bibfnamefont {Friedemann}\ \bibnamefont {Reinhard}}, \ and\ \bibinfo {author} {\bibfnamefont {Paola}\ \bibnamefont {Cappellaro}},\ }\bibfield  {title} {\enquote {\bibinfo {title} {Quantum sensing},}\ }\href@noop {} {\bibfield  {journal} {\bibinfo  {journal} {Reviews of modern physics}\ }\textbf {\bibinfo {volume} {89}},\ \bibinfo {pages} {035002} (\bibinfo {year} {2017})}\BibitemShut {NoStop}%
\bibitem [{\citenamefont {Giovannetti}\ \emph {et~al.}(2011)\citenamefont {Giovannetti}, \citenamefont {Lloyd},\ and\ \citenamefont {Maccone}}]{Giovannetti_2011}%
  \BibitemOpen
  \bibfield  {author} {\bibinfo {author} {\bibfnamefont {Vittorio}\ \bibnamefont {Giovannetti}}, \bibinfo {author} {\bibfnamefont {Seth}\ \bibnamefont {Lloyd}}, \ and\ \bibinfo {author} {\bibfnamefont {Lorenzo}\ \bibnamefont {Maccone}},\ }\bibfield  {title} {\enquote {\bibinfo {title} {Advances in quantum metrology},}\ }\href {\doibase 10.1038/nphoton.2011.35} {\bibfield  {journal} {\bibinfo  {journal} {Nature Photonics}\ }\textbf {\bibinfo {volume} {5}},\ \bibinfo {pages} {222–229} (\bibinfo {year} {2011})}\BibitemShut {NoStop}%
\bibitem [{\citenamefont {Greenberger}\ \emph {et~al.}(1989)\citenamefont {Greenberger}, \citenamefont {Horne},\ and\ \citenamefont {Zeilinger}}]{Greenberger1989}%
  \BibitemOpen
  \bibfield  {author} {\bibinfo {author} {\bibfnamefont {Daniel~M.}\ \bibnamefont {Greenberger}}, \bibinfo {author} {\bibfnamefont {Michael~A.}\ \bibnamefont {Horne}}, \ and\ \bibinfo {author} {\bibfnamefont {Anton}\ \bibnamefont {Zeilinger}},\ }\enquote {\bibinfo {title} {Going beyond bell's theorem},}\ in\ \href {\doibase 10.1007/978-94-017-0849-4_10} {\emph {\bibinfo {booktitle} {Bell's Theorem, Quantum Theory and Conceptions of the Universe}}},\ \bibinfo {editor} {edited by\ \bibinfo {editor} {\bibfnamefont {Menas}\ \bibnamefont {Kafatos}}}\ (\bibinfo  {publisher} {Springer Netherlands},\ \bibinfo {address} {Dordrecht},\ \bibinfo {year} {1989})\ pp.\ \bibinfo {pages} {69--72}\BibitemShut {NoStop}%
\bibitem [{\citenamefont {Huelga}\ \emph {et~al.}(1997)\citenamefont {Huelga}, \citenamefont {Macchiavello}, \citenamefont {Pellizzari}, \citenamefont {Ekert}, \citenamefont {Plenio},\ and\ \citenamefont {Cirac}}]{PhysRevLett.79.3865}%
  \BibitemOpen
  \bibfield  {author} {\bibinfo {author} {\bibfnamefont {S.~F.}\ \bibnamefont {Huelga}}, \bibinfo {author} {\bibfnamefont {C.}~\bibnamefont {Macchiavello}}, \bibinfo {author} {\bibfnamefont {T.}~\bibnamefont {Pellizzari}}, \bibinfo {author} {\bibfnamefont {A.~K.}\ \bibnamefont {Ekert}}, \bibinfo {author} {\bibfnamefont {M.~B.}\ \bibnamefont {Plenio}}, \ and\ \bibinfo {author} {\bibfnamefont {J.~I.}\ \bibnamefont {Cirac}},\ }\bibfield  {title} {\enquote {\bibinfo {title} {Improvement of frequency standards with quantum entanglement},}\ }\href {\doibase 10.1103/PhysRevLett.79.3865} {\bibfield  {journal} {\bibinfo  {journal} {Phys. Rev. Lett.}\ }\textbf {\bibinfo {volume} {79}},\ \bibinfo {pages} {3865--3868} (\bibinfo {year} {1997})}\BibitemShut {NoStop}%
\bibitem [{\citenamefont {Shaji}\ and\ \citenamefont {Caves}(2007)}]{shaji2007qubit}%
  \BibitemOpen
  \bibfield  {author} {\bibinfo {author} {\bibfnamefont {Anil}\ \bibnamefont {Shaji}}\ and\ \bibinfo {author} {\bibfnamefont {Carlton~M}\ \bibnamefont {Caves}},\ }\bibfield  {title} {\enquote {\bibinfo {title} {Qubit metrology and decoherence},}\ }\href@noop {} {\bibfield  {journal} {\bibinfo  {journal} {Physical Review A—Atomic, Molecular, and Optical Physics}\ }\textbf {\bibinfo {volume} {76}},\ \bibinfo {pages} {032111} (\bibinfo {year} {2007})}\BibitemShut {NoStop}%
\bibitem [{\citenamefont {Demkowicz-Dobrza{\'n}ski}\ \emph {et~al.}(2012)\citenamefont {Demkowicz-Dobrza{\'n}ski}, \citenamefont {Ko{\l}ody{\'n}ski},\ and\ \citenamefont {Gu{\c{t}}{\u{a}}}}]{demkowicz2012elusive}%
  \BibitemOpen
  \bibfield  {author} {\bibinfo {author} {\bibfnamefont {Rafa{\l}}\ \bibnamefont {Demkowicz-Dobrza{\'n}ski}}, \bibinfo {author} {\bibfnamefont {Jan}\ \bibnamefont {Ko{\l}ody{\'n}ski}}, \ and\ \bibinfo {author} {\bibfnamefont {M{\u{a}}d{\u{a}}lin}\ \bibnamefont {Gu{\c{t}}{\u{a}}}},\ }\bibfield  {title} {\enquote {\bibinfo {title} {The elusive heisenberg limit in quantum-enhanced metrology},}\ }\href@noop {} {\bibfield  {journal} {\bibinfo  {journal} {Nature communications}\ }\textbf {\bibinfo {volume} {3}},\ \bibinfo {pages} {1063} (\bibinfo {year} {2012})}\BibitemShut {NoStop}%
\bibitem [{\citenamefont {Brask}\ \emph {et~al.}(2015)\citenamefont {Brask}, \citenamefont {Chaves},\ and\ \citenamefont {Ko{\l}ody{\'n}ski}}]{brask2015improved}%
  \BibitemOpen
  \bibfield  {author} {\bibinfo {author} {\bibfnamefont {Jonatan~Bohr}\ \bibnamefont {Brask}}, \bibinfo {author} {\bibfnamefont {Rafael}\ \bibnamefont {Chaves}}, \ and\ \bibinfo {author} {\bibfnamefont {Janek}\ \bibnamefont {Ko{\l}ody{\'n}ski}},\ }\bibfield  {title} {\enquote {\bibinfo {title} {Improved quantum magnetometry beyond the standard quantum limit},}\ }\href@noop {} {\bibfield  {journal} {\bibinfo  {journal} {Physical Review X}\ }\textbf {\bibinfo {volume} {5}},\ \bibinfo {pages} {031010} (\bibinfo {year} {2015})}\BibitemShut {NoStop}%
\bibitem [{\citenamefont {Chaves}\ \emph {et~al.}(2013)\citenamefont {Chaves}, \citenamefont {Brask}, \citenamefont {Markiewicz}, \citenamefont {Ko{\l}ody{\'n}ski},\ and\ \citenamefont {Ac{\'\i}n}}]{chaves2013noisy}%
  \BibitemOpen
  \bibfield  {author} {\bibinfo {author} {\bibfnamefont {R}~\bibnamefont {Chaves}}, \bibinfo {author} {\bibfnamefont {JB}~\bibnamefont {Brask}}, \bibinfo {author} {\bibfnamefont {Marcin}\ \bibnamefont {Markiewicz}}, \bibinfo {author} {\bibfnamefont {J}~\bibnamefont {Ko{\l}ody{\'n}ski}}, \ and\ \bibinfo {author} {\bibfnamefont {A}~\bibnamefont {Ac{\'\i}n}},\ }\bibfield  {title} {\enquote {\bibinfo {title} {Noisy metrology beyond the standard quantum limit},}\ }\href@noop {} {\bibfield  {journal} {\bibinfo  {journal} {Physical review letters}\ }\textbf {\bibinfo {volume} {111}},\ \bibinfo {pages} {120401} (\bibinfo {year} {2013})}\BibitemShut {NoStop}%
\bibitem [{\citenamefont {Kessler}\ \emph {et~al.}(2014)\citenamefont {Kessler}, \citenamefont {Lovchinsky}, \citenamefont {Sushkov},\ and\ \citenamefont {Lukin}}]{GHZ_QEC1}%
  \BibitemOpen
  \bibfield  {author} {\bibinfo {author} {\bibfnamefont {E.~M.}\ \bibnamefont {Kessler}}, \bibinfo {author} {\bibfnamefont {I.}~\bibnamefont {Lovchinsky}}, \bibinfo {author} {\bibfnamefont {A.~O.}\ \bibnamefont {Sushkov}}, \ and\ \bibinfo {author} {\bibfnamefont {M.~D.}\ \bibnamefont {Lukin}},\ }\bibfield  {title} {\enquote {\bibinfo {title} {Quantum error correction for metrology},}\ }\href {\doibase 10.1103/PhysRevLett.112.150802} {\bibfield  {journal} {\bibinfo  {journal} {Phys. Rev. Lett.}\ }\textbf {\bibinfo {volume} {112}},\ \bibinfo {pages} {150802} (\bibinfo {year} {2014})}\BibitemShut {NoStop}%
\bibitem [{\citenamefont {D\"ur}\ \emph {et~al.}(2014)\citenamefont {D\"ur}, \citenamefont {Skotiniotis}, \citenamefont {Fr\"owis},\ and\ \citenamefont {Kraus}}]{GHZ_QEC2}%
  \BibitemOpen
  \bibfield  {author} {\bibinfo {author} {\bibfnamefont {W.}~\bibnamefont {D\"ur}}, \bibinfo {author} {\bibfnamefont {M.}~\bibnamefont {Skotiniotis}}, \bibinfo {author} {\bibfnamefont {F.}~\bibnamefont {Fr\"owis}}, \ and\ \bibinfo {author} {\bibfnamefont {B.}~\bibnamefont {Kraus}},\ }\bibfield  {title} {\enquote {\bibinfo {title} {Improved quantum metrology using quantum error correction},}\ }\href {\doibase 10.1103/PhysRevLett.112.080801} {\bibfield  {journal} {\bibinfo  {journal} {Phys. Rev. Lett.}\ }\textbf {\bibinfo {volume} {112}},\ \bibinfo {pages} {080801} (\bibinfo {year} {2014})}\BibitemShut {NoStop}%
\bibitem [{\citenamefont {Arrad}\ \emph {et~al.}(2014)\citenamefont {Arrad}, \citenamefont {Vinkler}, \citenamefont {Aharonov},\ and\ \citenamefont {Retzker}}]{PhysRevLett.112.150801}%
  \BibitemOpen
  \bibfield  {author} {\bibinfo {author} {\bibfnamefont {G.}~\bibnamefont {Arrad}}, \bibinfo {author} {\bibfnamefont {Y.}~\bibnamefont {Vinkler}}, \bibinfo {author} {\bibfnamefont {D.}~\bibnamefont {Aharonov}}, \ and\ \bibinfo {author} {\bibfnamefont {A.}~\bibnamefont {Retzker}},\ }\bibfield  {title} {\enquote {\bibinfo {title} {Increasing sensing resolution with error correction},}\ }\href {\doibase 10.1103/PhysRevLett.112.150801} {\bibfield  {journal} {\bibinfo  {journal} {Phys. Rev. Lett.}\ }\textbf {\bibinfo {volume} {112}},\ \bibinfo {pages} {150801} (\bibinfo {year} {2014})}\BibitemShut {NoStop}%
\bibitem [{\citenamefont {Isogawa}\ \emph {et~al.}(2023)\citenamefont {Isogawa}, \citenamefont {Matsuzaki},\ and\ \citenamefont {Ishi-Hayase}}]{isogawa2023vector}%
  \BibitemOpen
  \bibfield  {author} {\bibinfo {author} {\bibfnamefont {Takuya}\ \bibnamefont {Isogawa}}, \bibinfo {author} {\bibfnamefont {Yuichiro}\ \bibnamefont {Matsuzaki}}, \ and\ \bibinfo {author} {\bibfnamefont {Junko}\ \bibnamefont {Ishi-Hayase}},\ }\bibfield  {title} {\enquote {\bibinfo {title} {Vector dc magnetic-field sensing with a reference microwave field using perfectly aligned nitrogen-vacancy centers in diamond},}\ }\href@noop {} {\bibfield  {journal} {\bibinfo  {journal} {Physical Review A}\ }\textbf {\bibinfo {volume} {107}},\ \bibinfo {pages} {062423} (\bibinfo {year} {2023})}\BibitemShut {NoStop}%
\bibitem [{\citenamefont {Taylor}\ \emph {et~al.}(2008)\citenamefont {Taylor}, \citenamefont {Cappellaro}, \citenamefont {Childress}, \citenamefont {Jiang}, \citenamefont {Budker}, \citenamefont {Hemmer}, \citenamefont {Yacoby}, \citenamefont {Walsworth},\ and\ \citenamefont {Lukin}}]{taylor2008high}%
  \BibitemOpen
  \bibfield  {author} {\bibinfo {author} {\bibfnamefont {Jacob~M}\ \bibnamefont {Taylor}}, \bibinfo {author} {\bibfnamefont {Paola}\ \bibnamefont {Cappellaro}}, \bibinfo {author} {\bibfnamefont {Lilian}\ \bibnamefont {Childress}}, \bibinfo {author} {\bibfnamefont {Liang}\ \bibnamefont {Jiang}}, \bibinfo {author} {\bibfnamefont {Dmitry}\ \bibnamefont {Budker}}, \bibinfo {author} {\bibfnamefont {PR}~\bibnamefont {Hemmer}}, \bibinfo {author} {\bibfnamefont {Amir}\ \bibnamefont {Yacoby}}, \bibinfo {author} {\bibfnamefont {Ronald}\ \bibnamefont {Walsworth}}, \ and\ \bibinfo {author} {\bibfnamefont {MD}~\bibnamefont {Lukin}},\ }\bibfield  {title} {\enquote {\bibinfo {title} {High-sensitivity diamond magnetometer with nanoscale resolution},}\ }\href@noop {} {\bibfield  {journal} {\bibinfo  {journal} {Nature Physics}\ }\textbf {\bibinfo {volume} {4}},\ \bibinfo {pages} {810--816} (\bibinfo {year} {2008})}\BibitemShut {NoStop}%
\bibitem [{\citenamefont {Bal}\ \emph {et~al.}(2012)\citenamefont {Bal}, \citenamefont {Deng}, \citenamefont {Orgiazzi}, \citenamefont {Ong},\ and\ \citenamefont {Lupascu}}]{bal2012ultrasensitive}%
  \BibitemOpen
  \bibfield  {author} {\bibinfo {author} {\bibfnamefont {Mustafa}\ \bibnamefont {Bal}}, \bibinfo {author} {\bibfnamefont {Chunqing}\ \bibnamefont {Deng}}, \bibinfo {author} {\bibfnamefont {Jean-Luc}\ \bibnamefont {Orgiazzi}}, \bibinfo {author} {\bibfnamefont {FR}~\bibnamefont {Ong}}, \ and\ \bibinfo {author} {\bibfnamefont {Adrian}\ \bibnamefont {Lupascu}},\ }\bibfield  {title} {\enquote {\bibinfo {title} {Ultrasensitive magnetic field detection using a single artificial atom},}\ }\href@noop {} {\bibfield  {journal} {\bibinfo  {journal} {Nature communications}\ }\textbf {\bibinfo {volume} {3}},\ \bibinfo {pages} {1324} (\bibinfo {year} {2012})}\BibitemShut {NoStop}%
\bibitem [{\citenamefont {Bauch}\ \emph {et~al.}(2020)\citenamefont {Bauch}, \citenamefont {Singh}, \citenamefont {Lee}, \citenamefont {Hart}, \citenamefont {Schloss}, \citenamefont {Turner}, \citenamefont {Barry}, \citenamefont {Pham}, \citenamefont {Bar-Gill}, \citenamefont {Yelin} \emph {et~al.}}]{bauch2020decoherence}%
  \BibitemOpen
  \bibfield  {author} {\bibinfo {author} {\bibfnamefont {Erik}\ \bibnamefont {Bauch}}, \bibinfo {author} {\bibfnamefont {Swati}\ \bibnamefont {Singh}}, \bibinfo {author} {\bibfnamefont {Junghyun}\ \bibnamefont {Lee}}, \bibinfo {author} {\bibfnamefont {Connor~A}\ \bibnamefont {Hart}}, \bibinfo {author} {\bibfnamefont {Jennifer~M}\ \bibnamefont {Schloss}}, \bibinfo {author} {\bibfnamefont {Matthew~J}\ \bibnamefont {Turner}}, \bibinfo {author} {\bibfnamefont {John~F}\ \bibnamefont {Barry}}, \bibinfo {author} {\bibfnamefont {Linh~M}\ \bibnamefont {Pham}}, \bibinfo {author} {\bibfnamefont {Nir}\ \bibnamefont {Bar-Gill}}, \bibinfo {author} {\bibfnamefont {Susanne~F}\ \bibnamefont {Yelin}},  \emph {et~al.},\ }\bibfield  {title} {\enquote {\bibinfo {title} {Decoherence of ensembles of nitrogen-vacancy centers in diamond},}\ }\href@noop {} {\bibfield  {journal} {\bibinfo  {journal} {Physical Review B}\ }\textbf {\bibinfo {volume} {102}},\ \bibinfo {pages} {134210} (\bibinfo {year} {2020})}\BibitemShut {NoStop}%
\bibitem [{\citenamefont {Hayashi}\ \emph {et~al.}(2020)\citenamefont {Hayashi}, \citenamefont {Matsuzaki}, \citenamefont {Ashida}, \citenamefont {Onoda}, \citenamefont {Abe}, \citenamefont {Ohshima}, \citenamefont {Hatano}, \citenamefont {Taniguchi}, \citenamefont {Morishita}, \citenamefont {Fujiwara} \emph {et~al.}}]{hayashi2020experimental}%
  \BibitemOpen
  \bibfield  {author} {\bibinfo {author} {\bibfnamefont {Kan}\ \bibnamefont {Hayashi}}, \bibinfo {author} {\bibfnamefont {Yuichiro}\ \bibnamefont {Matsuzaki}}, \bibinfo {author} {\bibfnamefont {Takaki}\ \bibnamefont {Ashida}}, \bibinfo {author} {\bibfnamefont {Shinobu}\ \bibnamefont {Onoda}}, \bibinfo {author} {\bibfnamefont {Hiroshi}\ \bibnamefont {Abe}}, \bibinfo {author} {\bibfnamefont {Takeshi}\ \bibnamefont {Ohshima}}, \bibinfo {author} {\bibfnamefont {Mutsuko}\ \bibnamefont {Hatano}}, \bibinfo {author} {\bibfnamefont {Takashi}\ \bibnamefont {Taniguchi}}, \bibinfo {author} {\bibfnamefont {Hiroki}\ \bibnamefont {Morishita}}, \bibinfo {author} {\bibfnamefont {Masanori}\ \bibnamefont {Fujiwara}},  \emph {et~al.},\ }\bibfield  {title} {\enquote {\bibinfo {title} {Experimental and theoretical analysis of noise strength and environmental correlation time for ensembles of nitrogen-vacancy centers in diamond},}\ }\href@noop {} {\bibfield  {journal} {\bibinfo  {journal} {Journal of the Physical Society of Japan}\
  }\textbf {\bibinfo {volume} {89}},\ \bibinfo {pages} {054708} (\bibinfo {year} {2020})}\BibitemShut {NoStop}%
\bibitem [{\citenamefont {Palma}\ \emph {et~al.}(1996)\citenamefont {Palma}, \citenamefont {Suominen},\ and\ \citenamefont {Ekert}}]{palma1996quantum}%
  \BibitemOpen
  \bibfield  {author} {\bibinfo {author} {\bibfnamefont {G~Massimo}\ \bibnamefont {Palma}}, \bibinfo {author} {\bibfnamefont {Kalle-Antti}\ \bibnamefont {Suominen}}, \ and\ \bibinfo {author} {\bibfnamefont {Artur}\ \bibnamefont {Ekert}},\ }\bibfield  {title} {\enquote {\bibinfo {title} {Quantum computers and dissipation},}\ }\href@noop {} {\bibfield  {journal} {\bibinfo  {journal} {Proceedings of the Royal Society of London. Series A: Mathematical, Physical and Engineering Sciences}\ }\textbf {\bibinfo {volume} {452}},\ \bibinfo {pages} {567--584} (\bibinfo {year} {1996})}\BibitemShut {NoStop}%
\bibitem [{\citenamefont {Jeske}\ \emph {et~al.}(2014)\citenamefont {Jeske}, \citenamefont {Cole},\ and\ \citenamefont {Huelga}}]{GHZ_spatial}%
  \BibitemOpen
  \bibfield  {author} {\bibinfo {author} {\bibfnamefont {Jan}\ \bibnamefont {Jeske}}, \bibinfo {author} {\bibfnamefont {Jared~H}\ \bibnamefont {Cole}}, \ and\ \bibinfo {author} {\bibfnamefont {Susana~F}\ \bibnamefont {Huelga}},\ }\bibfield  {title} {\enquote {\bibinfo {title} {Quantum metrology subject to spatially correlated markovian noise: restoring the heisenberg limit},}\ }\href {\doibase 10.1088/1367-2630/16/7/073039} {\bibfield  {journal} {\bibinfo  {journal} {New Journal of Physics}\ }\textbf {\bibinfo {volume} {16}},\ \bibinfo {pages} {073039} (\bibinfo {year} {2014})}\BibitemShut {NoStop}%
\bibitem [{\citenamefont {Matsuzaki}\ \emph {et~al.}(2011)\citenamefont {Matsuzaki}, \citenamefont {Benjamin},\ and\ \citenamefont {Fitzsimons}}]{GHZ_nonMarkov1}%
  \BibitemOpen
  \bibfield  {author} {\bibinfo {author} {\bibfnamefont {Yuichiro}\ \bibnamefont {Matsuzaki}}, \bibinfo {author} {\bibfnamefont {Simon~C.}\ \bibnamefont {Benjamin}}, \ and\ \bibinfo {author} {\bibfnamefont {Joseph}\ \bibnamefont {Fitzsimons}},\ }\bibfield  {title} {\enquote {\bibinfo {title} {Magnetic field sensing beyond the standard quantum limit under the effect of decoherence},}\ }\href {\doibase 10.1103/PhysRevA.84.012103} {\bibfield  {journal} {\bibinfo  {journal} {Phys. Rev. A}\ }\textbf {\bibinfo {volume} {84}},\ \bibinfo {pages} {012103} (\bibinfo {year} {2011})}\BibitemShut {NoStop}%
\bibitem [{\citenamefont {Chin}\ \emph {et~al.}(2012)\citenamefont {Chin}, \citenamefont {Huelga},\ and\ \citenamefont {Plenio}}]{GHZ_nonMarkov2}%
  \BibitemOpen
  \bibfield  {author} {\bibinfo {author} {\bibfnamefont {Alex~W.}\ \bibnamefont {Chin}}, \bibinfo {author} {\bibfnamefont {Susana~F.}\ \bibnamefont {Huelga}}, \ and\ \bibinfo {author} {\bibfnamefont {Martin~B.}\ \bibnamefont {Plenio}},\ }\bibfield  {title} {\enquote {\bibinfo {title} {Quantum metrology in non-markovian environments},}\ }\href {\doibase 10.1103/PhysRevLett.109.233601} {\bibfield  {journal} {\bibinfo  {journal} {Phys. Rev. Lett.}\ }\textbf {\bibinfo {volume} {109}},\ \bibinfo {pages} {233601} (\bibinfo {year} {2012})}\BibitemShut {NoStop}%
\bibitem [{\citenamefont {Beau}\ and\ \citenamefont {del Campo}(2017)}]{GHZ_kbdyH_pbdyL}%
  \BibitemOpen
  \bibfield  {author} {\bibinfo {author} {\bibfnamefont {M.}~\bibnamefont {Beau}}\ and\ \bibinfo {author} {\bibfnamefont {A.}~\bibnamefont {del Campo}},\ }\bibfield  {title} {\enquote {\bibinfo {title} {Nonlinear quantum metrology of many-body open systems},}\ }\href {\doibase 10.1103/PhysRevLett.119.010403} {\bibfield  {journal} {\bibinfo  {journal} {Phys. Rev. Lett.}\ }\textbf {\bibinfo {volume} {119}},\ \bibinfo {pages} {010403} (\bibinfo {year} {2017})}\BibitemShut {NoStop}%
\bibitem [{\citenamefont {Matsuzaki}\ \emph {et~al.}(2018)\citenamefont {Matsuzaki}, \citenamefont {Saito},\ and\ \citenamefont {Munro}}]{matsuzaki2018quantum}%
  \BibitemOpen
  \bibfield  {author} {\bibinfo {author} {\bibfnamefont {Yuichiro}\ \bibnamefont {Matsuzaki}}, \bibinfo {author} {\bibfnamefont {Shiro}\ \bibnamefont {Saito}}, \ and\ \bibinfo {author} {\bibfnamefont {William~J.}\ \bibnamefont {Munro}},\ }\href@noop {} {\enquote {\bibinfo {title} {Quantum metrology at the heisenberg limit with the presence of independent dephasing},}\ } (\bibinfo {year} {2018}),\ \Eprint {http://arxiv.org/abs/1809.00176} {arXiv:1809.00176} \BibitemShut {NoStop}%
\bibitem [{\citenamefont {Kukita}\ \emph {et~al.}(2021)\citenamefont {Kukita}, \citenamefont {Matsuzaki},\ and\ \citenamefont {Kondo}}]{PhysRevApplied.16.064026}%
  \BibitemOpen
  \bibfield  {author} {\bibinfo {author} {\bibfnamefont {Shingo}\ \bibnamefont {Kukita}}, \bibinfo {author} {\bibfnamefont {Yuichiro}\ \bibnamefont {Matsuzaki}}, \ and\ \bibinfo {author} {\bibfnamefont {Yasushi}\ \bibnamefont {Kondo}},\ }\bibfield  {title} {\enquote {\bibinfo {title} {Heisenberg-limited quantum metrology using collective dephasing},}\ }\href {\doibase 10.1103/PhysRevApplied.16.064026} {\bibfield  {journal} {\bibinfo  {journal} {Phys. Rev. Appl.}\ }\textbf {\bibinfo {volume} {16}},\ \bibinfo {pages} {064026} (\bibinfo {year} {2021})}\BibitemShut {NoStop}%
\bibitem [{\citenamefont {Balasubramanian}\ \emph {et~al.}(2008)\citenamefont {Balasubramanian}, \citenamefont {Chan}, \citenamefont {Kolesov}, \citenamefont {Al-Hmoud}, \citenamefont {Tisler}, \citenamefont {Shin}, \citenamefont {Kim}, \citenamefont {Wojcik}, \citenamefont {Hemmer}, \citenamefont {Krueger} \emph {et~al.}}]{balasubramanian2008nanoscale}%
  \BibitemOpen
  \bibfield  {author} {\bibinfo {author} {\bibfnamefont {Gopalakrishnan}\ \bibnamefont {Balasubramanian}}, \bibinfo {author} {\bibfnamefont {IY}~\bibnamefont {Chan}}, \bibinfo {author} {\bibfnamefont {Roman}\ \bibnamefont {Kolesov}}, \bibinfo {author} {\bibfnamefont {Mohannad}\ \bibnamefont {Al-Hmoud}}, \bibinfo {author} {\bibfnamefont {Julia}\ \bibnamefont {Tisler}}, \bibinfo {author} {\bibfnamefont {Chang}\ \bibnamefont {Shin}}, \bibinfo {author} {\bibfnamefont {Changdong}\ \bibnamefont {Kim}}, \bibinfo {author} {\bibfnamefont {Aleksander}\ \bibnamefont {Wojcik}}, \bibinfo {author} {\bibfnamefont {Philip~R}\ \bibnamefont {Hemmer}}, \bibinfo {author} {\bibfnamefont {Anke}\ \bibnamefont {Krueger}},  \emph {et~al.},\ }\bibfield  {title} {\enquote {\bibinfo {title} {Nanoscale imaging magnetometry with diamond spins under ambient conditions},}\ }\href@noop {} {\bibfield  {journal} {\bibinfo  {journal} {Nature}\ }\textbf {\bibinfo {volume} {455}},\ \bibinfo {pages} {648--651} (\bibinfo {year} {2008})}\BibitemShut
  {NoStop}%
\bibitem [{\citenamefont {Maze}\ \emph {et~al.}(2008)\citenamefont {Maze}, \citenamefont {Stanwix}, \citenamefont {Hodges}, \citenamefont {Hong}, \citenamefont {Taylor}, \citenamefont {Cappellaro}, \citenamefont {Jiang}, \citenamefont {Dutt}, \citenamefont {Togan}, \citenamefont {Zibrov} \emph {et~al.}}]{maze2008nanoscale}%
  \BibitemOpen
  \bibfield  {author} {\bibinfo {author} {\bibfnamefont {Jeronimo~R}\ \bibnamefont {Maze}}, \bibinfo {author} {\bibfnamefont {Paul~L}\ \bibnamefont {Stanwix}}, \bibinfo {author} {\bibfnamefont {James~S}\ \bibnamefont {Hodges}}, \bibinfo {author} {\bibfnamefont {Seungpyo}\ \bibnamefont {Hong}}, \bibinfo {author} {\bibfnamefont {Jacob~M}\ \bibnamefont {Taylor}}, \bibinfo {author} {\bibfnamefont {Paola}\ \bibnamefont {Cappellaro}}, \bibinfo {author} {\bibfnamefont {Liang}\ \bibnamefont {Jiang}}, \bibinfo {author} {\bibfnamefont {MV~Gurudev}\ \bibnamefont {Dutt}}, \bibinfo {author} {\bibfnamefont {Emre}\ \bibnamefont {Togan}}, \bibinfo {author} {\bibfnamefont {AS}~\bibnamefont {Zibrov}},  \emph {et~al.},\ }\bibfield  {title} {\enquote {\bibinfo {title} {Nanoscale magnetic sensing with an individual electronic spin in diamond},}\ }\href@noop {} {\bibfield  {journal} {\bibinfo  {journal} {Nature}\ }\textbf {\bibinfo {volume} {455}},\ \bibinfo {pages} {644--647} (\bibinfo {year} {2008})}\BibitemShut {NoStop}%
\bibitem [{\citenamefont {De~Lange}\ \emph {et~al.}(2010)\citenamefont {De~Lange}, \citenamefont {Wang}, \citenamefont {Riste}, \citenamefont {Dobrovitski},\ and\ \citenamefont {Hanson}}]{de2010universal}%
  \BibitemOpen
  \bibfield  {author} {\bibinfo {author} {\bibfnamefont {G}~\bibnamefont {De~Lange}}, \bibinfo {author} {\bibfnamefont {ZH}~\bibnamefont {Wang}}, \bibinfo {author} {\bibfnamefont {D}~\bibnamefont {Riste}}, \bibinfo {author} {\bibfnamefont {VV}~\bibnamefont {Dobrovitski}}, \ and\ \bibinfo {author} {\bibfnamefont {R}~\bibnamefont {Hanson}},\ }\bibfield  {title} {\enquote {\bibinfo {title} {Universal dynamical decoupling of a single solid-state spin from a spin bath},}\ }\href@noop {} {\bibfield  {journal} {\bibinfo  {journal} {Science}\ }\textbf {\bibinfo {volume} {330}},\ \bibinfo {pages} {60--63} (\bibinfo {year} {2010})}\BibitemShut {NoStop}%
\bibitem [{\citenamefont {Bylander}\ \emph {et~al.}(2011)\citenamefont {Bylander}, \citenamefont {Gustavsson}, \citenamefont {Yan}, \citenamefont {Yoshihara}, \citenamefont {Harrabi}, \citenamefont {Fitch}, \citenamefont {Cory}, \citenamefont {Nakamura}, \citenamefont {Tsai},\ and\ \citenamefont {Oliver}}]{bylander2011noise}%
  \BibitemOpen
  \bibfield  {author} {\bibinfo {author} {\bibfnamefont {Jonas}\ \bibnamefont {Bylander}}, \bibinfo {author} {\bibfnamefont {Simon}\ \bibnamefont {Gustavsson}}, \bibinfo {author} {\bibfnamefont {Fei}\ \bibnamefont {Yan}}, \bibinfo {author} {\bibfnamefont {Fumiki}\ \bibnamefont {Yoshihara}}, \bibinfo {author} {\bibfnamefont {Khalil}\ \bibnamefont {Harrabi}}, \bibinfo {author} {\bibfnamefont {George}\ \bibnamefont {Fitch}}, \bibinfo {author} {\bibfnamefont {David~G}\ \bibnamefont {Cory}}, \bibinfo {author} {\bibfnamefont {Yasunobu}\ \bibnamefont {Nakamura}}, \bibinfo {author} {\bibfnamefont {Jaw-Shen}\ \bibnamefont {Tsai}}, \ and\ \bibinfo {author} {\bibfnamefont {William~D}\ \bibnamefont {Oliver}},\ }\bibfield  {title} {\enquote {\bibinfo {title} {Noise spectroscopy through dynamical decoupling with a superconducting flux qubit},}\ }\href@noop {} {\bibfield  {journal} {\bibinfo  {journal} {Nature Physics}\ }\textbf {\bibinfo {volume} {7}},\ \bibinfo {pages} {565--570} (\bibinfo {year} {2011})}\BibitemShut
  {NoStop}%
\bibitem [{\citenamefont {Yoshihara}\ \emph {et~al.}(2014)\citenamefont {Yoshihara}, \citenamefont {Nakamura}, \citenamefont {Yan}, \citenamefont {Gustavsson}, \citenamefont {Bylander}, \citenamefont {Oliver},\ and\ \citenamefont {Tsai}}]{yoshihara2014flux}%
  \BibitemOpen
  \bibfield  {author} {\bibinfo {author} {\bibfnamefont {Fumiki}\ \bibnamefont {Yoshihara}}, \bibinfo {author} {\bibfnamefont {Yasunobu}\ \bibnamefont {Nakamura}}, \bibinfo {author} {\bibfnamefont {Fei}\ \bibnamefont {Yan}}, \bibinfo {author} {\bibfnamefont {Simon}\ \bibnamefont {Gustavsson}}, \bibinfo {author} {\bibfnamefont {Jonas}\ \bibnamefont {Bylander}}, \bibinfo {author} {\bibfnamefont {William~D}\ \bibnamefont {Oliver}}, \ and\ \bibinfo {author} {\bibfnamefont {Jaw-Shen}\ \bibnamefont {Tsai}},\ }\bibfield  {title} {\enquote {\bibinfo {title} {Flux qubit noise spectroscopy using rabi oscillations under strong driving conditions},}\ }\href@noop {} {\bibfield  {journal} {\bibinfo  {journal} {Physical Review B}\ }\textbf {\bibinfo {volume} {89}},\ \bibinfo {pages} {020503} (\bibinfo {year} {2014})}\BibitemShut {NoStop}%
\bibitem [{\citenamefont {Appel}\ \emph {et~al.}(2015)\citenamefont {Appel}, \citenamefont {Ganzhorn}, \citenamefont {Neu},\ and\ \citenamefont {Maletinsky}}]{appel2015nanoscale}%
  \BibitemOpen
  \bibfield  {author} {\bibinfo {author} {\bibfnamefont {Patrick}\ \bibnamefont {Appel}}, \bibinfo {author} {\bibfnamefont {Marc}\ \bibnamefont {Ganzhorn}}, \bibinfo {author} {\bibfnamefont {Elke}\ \bibnamefont {Neu}}, \ and\ \bibinfo {author} {\bibfnamefont {Patrick}\ \bibnamefont {Maletinsky}},\ }\bibfield  {title} {\enquote {\bibinfo {title} {Nanoscale microwave imaging with a single electron spin in diamond},}\ }\href@noop {} {\bibfield  {journal} {\bibinfo  {journal} {New Journal of Physics}\ }\textbf {\bibinfo {volume} {17}},\ \bibinfo {pages} {112001} (\bibinfo {year} {2015})}\BibitemShut {NoStop}%
\bibitem [{\citenamefont {Cai}\ \emph {et~al.}(2012)\citenamefont {Cai}, \citenamefont {Naydenov}, \citenamefont {Pfeiffer}, \citenamefont {McGuinness}, \citenamefont {Jahnke}, \citenamefont {Jelezko}, \citenamefont {Plenio},\ and\ \citenamefont {Retzker}}]{cai2012robust}%
  \BibitemOpen
  \bibfield  {author} {\bibinfo {author} {\bibfnamefont {JM}~\bibnamefont {Cai}}, \bibinfo {author} {\bibfnamefont {Boris}\ \bibnamefont {Naydenov}}, \bibinfo {author} {\bibfnamefont {Rainer}\ \bibnamefont {Pfeiffer}}, \bibinfo {author} {\bibfnamefont {Liam~P}\ \bibnamefont {McGuinness}}, \bibinfo {author} {\bibfnamefont {Kay~D}\ \bibnamefont {Jahnke}}, \bibinfo {author} {\bibfnamefont {Fedor}\ \bibnamefont {Jelezko}}, \bibinfo {author} {\bibfnamefont {Martin~B}\ \bibnamefont {Plenio}}, \ and\ \bibinfo {author} {\bibfnamefont {Alex}\ \bibnamefont {Retzker}},\ }\bibfield  {title} {\enquote {\bibinfo {title} {Robust dynamical decoupling with concatenated continuous driving},}\ }\href@noop {} {\bibfield  {journal} {\bibinfo  {journal} {New Journal of Physics}\ }\textbf {\bibinfo {volume} {14}},\ \bibinfo {pages} {113023} (\bibinfo {year} {2012})}\BibitemShut {NoStop}%
\bibitem [{\citenamefont {Okaniwa}\ \emph {et~al.}(2024)\citenamefont {Okaniwa}, \citenamefont {Mikawa}, \citenamefont {Matsuzaki}, \citenamefont {Yamaguchi}, \citenamefont {Suzuki}, \citenamefont {Tokuda}, \citenamefont {Watanabe}, \citenamefont {Mizuochi}, \citenamefont {Sasaki}, \citenamefont {Kobayashi} \emph {et~al.}}]{okaniwa2024frequency}%
  \BibitemOpen
  \bibfield  {author} {\bibinfo {author} {\bibfnamefont {Ryusei}\ \bibnamefont {Okaniwa}}, \bibinfo {author} {\bibfnamefont {Takumi}\ \bibnamefont {Mikawa}}, \bibinfo {author} {\bibfnamefont {Yuichiro}\ \bibnamefont {Matsuzaki}}, \bibinfo {author} {\bibfnamefont {Tatsuma}\ \bibnamefont {Yamaguchi}}, \bibinfo {author} {\bibfnamefont {Rui}\ \bibnamefont {Suzuki}}, \bibinfo {author} {\bibfnamefont {Norio}\ \bibnamefont {Tokuda}}, \bibinfo {author} {\bibfnamefont {Hideyuki}\ \bibnamefont {Watanabe}}, \bibinfo {author} {\bibfnamefont {Norikazu}\ \bibnamefont {Mizuochi}}, \bibinfo {author} {\bibfnamefont {Kento}\ \bibnamefont {Sasaki}}, \bibinfo {author} {\bibfnamefont {Kensuke}\ \bibnamefont {Kobayashi}},  \emph {et~al.},\ }\bibfield  {title} {\enquote {\bibinfo {title} {Frequency-tunable magnetic field sensing using continuous-wave optically detected magnetic resonance with nitrogen-vacancy centers in diamond},}\ }\href@noop {} {\bibfield  {journal} {\bibinfo  {journal} {Journal of Applied Physics}\ }\textbf
  {\bibinfo {volume} {135}} (\bibinfo {year} {2024})}\BibitemShut {NoStop}%
\bibitem [{\citenamefont {Genov}\ \emph {et~al.}(2020)\citenamefont {Genov}, \citenamefont {Ben-Shalom}, \citenamefont {Jelezko}, \citenamefont {Retzker},\ and\ \citenamefont {Bar-Gill}}]{PhysRevResearch.2.033216}%
  \BibitemOpen
  \bibfield  {author} {\bibinfo {author} {\bibfnamefont {Genko~T.}\ \bibnamefont {Genov}}, \bibinfo {author} {\bibfnamefont {Yachel}\ \bibnamefont {Ben-Shalom}}, \bibinfo {author} {\bibfnamefont {Fedor}\ \bibnamefont {Jelezko}}, \bibinfo {author} {\bibfnamefont {Alex}\ \bibnamefont {Retzker}}, \ and\ \bibinfo {author} {\bibfnamefont {Nir}\ \bibnamefont {Bar-Gill}},\ }\bibfield  {title} {\enquote {\bibinfo {title} {Efficient and robust signal sensing by sequences of adiabatic chirped pulses},}\ }\href {\doibase 10.1103/PhysRevResearch.2.033216} {\bibfield  {journal} {\bibinfo  {journal} {Phys. Rev. Res.}\ }\textbf {\bibinfo {volume} {2}},\ \bibinfo {pages} {033216} (\bibinfo {year} {2020})}\BibitemShut {NoStop}%
\bibitem [{\citenamefont {Tian}\ \emph {et~al.}(2020)\citenamefont {Tian}, \citenamefont {Liu}, \citenamefont {Liu}, \citenamefont {Yang}, \citenamefont {Betzholz}, \citenamefont {Said}, \citenamefont {Jelezko},\ and\ \citenamefont {Cai}}]{PhysRevA.102.043707}%
  \BibitemOpen
  \bibfield  {author} {\bibinfo {author} {\bibfnamefont {Jiazhao}\ \bibnamefont {Tian}}, \bibinfo {author} {\bibfnamefont {Haibin}\ \bibnamefont {Liu}}, \bibinfo {author} {\bibfnamefont {Yu}~\bibnamefont {Liu}}, \bibinfo {author} {\bibfnamefont {Pengcheng}\ \bibnamefont {Yang}}, \bibinfo {author} {\bibfnamefont {Ralf}\ \bibnamefont {Betzholz}}, \bibinfo {author} {\bibfnamefont {Ressa~S.}\ \bibnamefont {Said}}, \bibinfo {author} {\bibfnamefont {Fedor}\ \bibnamefont {Jelezko}}, \ and\ \bibinfo {author} {\bibfnamefont {Jianming}\ \bibnamefont {Cai}},\ }\bibfield  {title} {\enquote {\bibinfo {title} {Quantum optimal control using phase-modulated driving fields},}\ }\href {\doibase 10.1103/PhysRevA.102.043707} {\bibfield  {journal} {\bibinfo  {journal} {Phys. Rev. A}\ }\textbf {\bibinfo {volume} {102}},\ \bibinfo {pages} {043707} (\bibinfo {year} {2020})}\BibitemShut {NoStop}%
\bibitem [{\citenamefont {Tyryshkin}\ \emph {et~al.}(2012)\citenamefont {Tyryshkin}, \citenamefont {Tojo}, \citenamefont {Morton}, \citenamefont {Riemann}, \citenamefont {Abrosimov}, \citenamefont {Becker}, \citenamefont {Pohl}, \citenamefont {Schenkel}, \citenamefont {Thewalt}, \citenamefont {Itoh} \emph {et~al.}}]{tyryshkin2012electron}%
  \BibitemOpen
  \bibfield  {author} {\bibinfo {author} {\bibfnamefont {Alexei~M}\ \bibnamefont {Tyryshkin}}, \bibinfo {author} {\bibfnamefont {Shinichi}\ \bibnamefont {Tojo}}, \bibinfo {author} {\bibfnamefont {John~JL}\ \bibnamefont {Morton}}, \bibinfo {author} {\bibfnamefont {Helge}\ \bibnamefont {Riemann}}, \bibinfo {author} {\bibfnamefont {Nikolai~V}\ \bibnamefont {Abrosimov}}, \bibinfo {author} {\bibfnamefont {Peter}\ \bibnamefont {Becker}}, \bibinfo {author} {\bibfnamefont {Hans-Joachim}\ \bibnamefont {Pohl}}, \bibinfo {author} {\bibfnamefont {Thomas}\ \bibnamefont {Schenkel}}, \bibinfo {author} {\bibfnamefont {Michael~LW}\ \bibnamefont {Thewalt}}, \bibinfo {author} {\bibfnamefont {Kohei~M}\ \bibnamefont {Itoh}},  \emph {et~al.},\ }\bibfield  {title} {\enquote {\bibinfo {title} {Electron spin coherence exceeding seconds in high-purity silicon},}\ }\href@noop {} {\bibfield  {journal} {\bibinfo  {journal} {Nature materials}\ }\textbf {\bibinfo {volume} {11}},\ \bibinfo {pages} {143--147} (\bibinfo {year}
  {2012})}\BibitemShut {NoStop}%
\bibitem [{\citenamefont {Wolf}\ \emph {et~al.}(2015)\citenamefont {Wolf}, \citenamefont {Neumann}, \citenamefont {Nakamura}, \citenamefont {Sumiya}, \citenamefont {Ohshima}, \citenamefont {Isoya},\ and\ \citenamefont {Wrachtrup}}]{wolf2015subpicotesla}%
  \BibitemOpen
  \bibfield  {author} {\bibinfo {author} {\bibfnamefont {Thomas}\ \bibnamefont {Wolf}}, \bibinfo {author} {\bibfnamefont {Philipp}\ \bibnamefont {Neumann}}, \bibinfo {author} {\bibfnamefont {Kazuo}\ \bibnamefont {Nakamura}}, \bibinfo {author} {\bibfnamefont {Hitoshi}\ \bibnamefont {Sumiya}}, \bibinfo {author} {\bibfnamefont {Takeshi}\ \bibnamefont {Ohshima}}, \bibinfo {author} {\bibfnamefont {Junichi}\ \bibnamefont {Isoya}}, \ and\ \bibinfo {author} {\bibfnamefont {J{\"o}rg}\ \bibnamefont {Wrachtrup}},\ }\bibfield  {title} {\enquote {\bibinfo {title} {Subpicotesla diamond magnetometry},}\ }\href@noop {} {\bibfield  {journal} {\bibinfo  {journal} {Physical Review X}\ }\textbf {\bibinfo {volume} {5}},\ \bibinfo {pages} {041001} (\bibinfo {year} {2015})}\BibitemShut {NoStop}%
\bibitem [{\citenamefont {Zhu}\ \emph {et~al.}(2011)\citenamefont {Zhu}, \citenamefont {Saito}, \citenamefont {Kemp}, \citenamefont {Kakuyanagi}, \citenamefont {Karimoto}, \citenamefont {Nakano}, \citenamefont {Munro}, \citenamefont {Tokura}, \citenamefont {Everitt}, \citenamefont {Nemoto} \emph {et~al.}}]{zhu2011coherent}%
  \BibitemOpen
  \bibfield  {author} {\bibinfo {author} {\bibfnamefont {Xiaobo}\ \bibnamefont {Zhu}}, \bibinfo {author} {\bibfnamefont {Shiro}\ \bibnamefont {Saito}}, \bibinfo {author} {\bibfnamefont {Alexander}\ \bibnamefont {Kemp}}, \bibinfo {author} {\bibfnamefont {Kosuke}\ \bibnamefont {Kakuyanagi}}, \bibinfo {author} {\bibfnamefont {Shin-ichi}\ \bibnamefont {Karimoto}}, \bibinfo {author} {\bibfnamefont {Hayato}\ \bibnamefont {Nakano}}, \bibinfo {author} {\bibfnamefont {William~J}\ \bibnamefont {Munro}}, \bibinfo {author} {\bibfnamefont {Yasuhiro}\ \bibnamefont {Tokura}}, \bibinfo {author} {\bibfnamefont {Mark~S}\ \bibnamefont {Everitt}}, \bibinfo {author} {\bibfnamefont {Kae}\ \bibnamefont {Nemoto}},  \emph {et~al.},\ }\bibfield  {title} {\enquote {\bibinfo {title} {Coherent coupling of a superconducting flux qubit to an electron spin ensemble in diamond},}\ }\href@noop {} {\bibfield  {journal} {\bibinfo  {journal} {Nature}\ }\textbf {\bibinfo {volume} {478}},\ \bibinfo {pages} {221--224} (\bibinfo {year}
  {2011})}\BibitemShut {NoStop}%
\bibitem [{\citenamefont {Kubo}\ \emph {et~al.}(2011)\citenamefont {Kubo}, \citenamefont {Grezes}, \citenamefont {Dewes}, \citenamefont {Umeda}, \citenamefont {Isoya}, \citenamefont {Sumiya}, \citenamefont {Morishita}, \citenamefont {Abe}, \citenamefont {Onoda}, \citenamefont {Ohshima} \emph {et~al.}}]{kubo2011hybrid}%
  \BibitemOpen
  \bibfield  {author} {\bibinfo {author} {\bibfnamefont {Yuimaru}\ \bibnamefont {Kubo}}, \bibinfo {author} {\bibfnamefont {Cecile}\ \bibnamefont {Grezes}}, \bibinfo {author} {\bibfnamefont {Andreas}\ \bibnamefont {Dewes}}, \bibinfo {author} {\bibfnamefont {T}~\bibnamefont {Umeda}}, \bibinfo {author} {\bibfnamefont {Junichi}\ \bibnamefont {Isoya}}, \bibinfo {author} {\bibfnamefont {H}~\bibnamefont {Sumiya}}, \bibinfo {author} {\bibfnamefont {N}~\bibnamefont {Morishita}}, \bibinfo {author} {\bibfnamefont {H}~\bibnamefont {Abe}}, \bibinfo {author} {\bibfnamefont {S}~\bibnamefont {Onoda}}, \bibinfo {author} {\bibfnamefont {T}~\bibnamefont {Ohshima}},  \emph {et~al.},\ }\bibfield  {title} {\enquote {\bibinfo {title} {Hybrid quantum circuit with a superconducting qubit coupled to a spin ensemble},}\ }\href@noop {} {\bibfield  {journal} {\bibinfo  {journal} {Physical review letters}\ }\textbf {\bibinfo {volume} {107}},\ \bibinfo {pages} {220501} (\bibinfo {year} {2011})}\BibitemShut {NoStop}%
\bibitem [{\citenamefont {Zhu}\ \emph {et~al.}(2014)\citenamefont {Zhu}, \citenamefont {Matsuzaki}, \citenamefont {Ams{\"u}ss}, \citenamefont {Kakuyanagi}, \citenamefont {Shimo-Oka}, \citenamefont {Mizuochi}, \citenamefont {Nemoto}, \citenamefont {Semba}, \citenamefont {Munro},\ and\ \citenamefont {Saito}}]{zhu2014observation}%
  \BibitemOpen
  \bibfield  {author} {\bibinfo {author} {\bibfnamefont {Xiaobo}\ \bibnamefont {Zhu}}, \bibinfo {author} {\bibfnamefont {Yuichiro}\ \bibnamefont {Matsuzaki}}, \bibinfo {author} {\bibfnamefont {Robert}\ \bibnamefont {Ams{\"u}ss}}, \bibinfo {author} {\bibfnamefont {Kosuke}\ \bibnamefont {Kakuyanagi}}, \bibinfo {author} {\bibfnamefont {Takaaki}\ \bibnamefont {Shimo-Oka}}, \bibinfo {author} {\bibfnamefont {Norikazu}\ \bibnamefont {Mizuochi}}, \bibinfo {author} {\bibfnamefont {Kae}\ \bibnamefont {Nemoto}}, \bibinfo {author} {\bibfnamefont {Kouichi}\ \bibnamefont {Semba}}, \bibinfo {author} {\bibfnamefont {William~J}\ \bibnamefont {Munro}}, \ and\ \bibinfo {author} {\bibfnamefont {Shiro}\ \bibnamefont {Saito}},\ }\bibfield  {title} {\enquote {\bibinfo {title} {Observation of dark states in a superconductor diamond quantum hybrid system},}\ }\href@noop {} {\bibfield  {journal} {\bibinfo  {journal} {Nature communications}\ }\textbf {\bibinfo {volume} {5}},\ \bibinfo {pages} {3524} (\bibinfo {year}
  {2014})}\BibitemShut {NoStop}%
\bibitem [{\citenamefont {Tanaka}\ \emph {et~al.}(2015)\citenamefont {Tanaka}, \citenamefont {Knott}, \citenamefont {Matsuzaki}, \citenamefont {Dooley}, \citenamefont {Yamaguchi}, \citenamefont {Munro},\ and\ \citenamefont {Saito}}]{tanaka2015proposed}%
  \BibitemOpen
  \bibfield  {author} {\bibinfo {author} {\bibfnamefont {Tohru}\ \bibnamefont {Tanaka}}, \bibinfo {author} {\bibfnamefont {Paul}\ \bibnamefont {Knott}}, \bibinfo {author} {\bibfnamefont {Yuichiro}\ \bibnamefont {Matsuzaki}}, \bibinfo {author} {\bibfnamefont {Shane}\ \bibnamefont {Dooley}}, \bibinfo {author} {\bibfnamefont {Hiroshi}\ \bibnamefont {Yamaguchi}}, \bibinfo {author} {\bibfnamefont {William~J}\ \bibnamefont {Munro}}, \ and\ \bibinfo {author} {\bibfnamefont {Shiro}\ \bibnamefont {Saito}},\ }\bibfield  {title} {\enquote {\bibinfo {title} {Proposed robust entanglement-based magnetic field sensor beyond the standard quantum limit},}\ }\href@noop {} {\bibfield  {journal} {\bibinfo  {journal} {Physical review letters}\ }\textbf {\bibinfo {volume} {115}},\ \bibinfo {pages} {170801} (\bibinfo {year} {2015})}\BibitemShut {NoStop}%
\bibitem [{\citenamefont {Dooley}\ \emph {et~al.}(2016)\citenamefont {Dooley}, \citenamefont {Yukawa}, \citenamefont {Matsuzaki}, \citenamefont {Knee}, \citenamefont {Munro},\ and\ \citenamefont {Nemoto}}]{dooley2016hybrid}%
  \BibitemOpen
  \bibfield  {author} {\bibinfo {author} {\bibfnamefont {Shane}\ \bibnamefont {Dooley}}, \bibinfo {author} {\bibfnamefont {Emi}\ \bibnamefont {Yukawa}}, \bibinfo {author} {\bibfnamefont {Yuichiro}\ \bibnamefont {Matsuzaki}}, \bibinfo {author} {\bibfnamefont {George~C}\ \bibnamefont {Knee}}, \bibinfo {author} {\bibfnamefont {William~J}\ \bibnamefont {Munro}}, \ and\ \bibinfo {author} {\bibfnamefont {Kae}\ \bibnamefont {Nemoto}},\ }\bibfield  {title} {\enquote {\bibinfo {title} {A hybrid-systems approach to spin squeezing using a highly dissipative ancillary system},}\ }\href@noop {} {\bibfield  {journal} {\bibinfo  {journal} {New Journal of Physics}\ }\textbf {\bibinfo {volume} {18}},\ \bibinfo {pages} {053011} (\bibinfo {year} {2016})}\BibitemShut {NoStop}%
\bibitem [{\citenamefont {Tatsuta}\ \emph {et~al.}(2024)\citenamefont {Tatsuta}, \citenamefont {Matsuzaki}, \citenamefont {Kuji},\ and\ \citenamefont {Shimizu}}]{tatsuta2024generation}%
  \BibitemOpen
  \bibfield  {author} {\bibinfo {author} {\bibfnamefont {Mamiko}\ \bibnamefont {Tatsuta}}, \bibinfo {author} {\bibfnamefont {Yuichiro}\ \bibnamefont {Matsuzaki}}, \bibinfo {author} {\bibfnamefont {Hiroki}\ \bibnamefont {Kuji}}, \ and\ \bibinfo {author} {\bibfnamefont {Akira}\ \bibnamefont {Shimizu}},\ }\bibfield  {title} {\enquote {\bibinfo {title} {Generation of a metrologically useful cat state through repetitive measurements},}\ }\href@noop {} {\bibfield  {journal} {\bibinfo  {journal} {arXiv preprint arXiv:2407.06829}\ } (\bibinfo {year} {2024})}\BibitemShut {NoStop}%
\bibitem [{\citenamefont {Matsuzaki}\ \emph {et~al.}(2022)\citenamefont {Matsuzaki}, \citenamefont {Imoto},\ and\ \citenamefont {Susa}}]{matsuzaki2022generation}%
  \BibitemOpen
  \bibfield  {author} {\bibinfo {author} {\bibfnamefont {Yuichiro}\ \bibnamefont {Matsuzaki}}, \bibinfo {author} {\bibfnamefont {Takashi}\ \bibnamefont {Imoto}}, \ and\ \bibinfo {author} {\bibfnamefont {Yuki}\ \bibnamefont {Susa}},\ }\bibfield  {title} {\enquote {\bibinfo {title} {Generation of multipartite entanglement between spin-1 particles with bifurcation-based quantum annealing},}\ }\href@noop {} {\bibfield  {journal} {\bibinfo  {journal} {Scientific Reports}\ }\textbf {\bibinfo {volume} {12}},\ \bibinfo {pages} {14964} (\bibinfo {year} {2022})}\BibitemShut {NoStop}%
\bibitem [{\citenamefont {Chou}\ \emph {et~al.}(2023)\citenamefont {Chou} \emph {et~al.}}]{Chou:2023hcc}%
  \BibitemOpen
  \bibfield  {author} {\bibinfo {author} {\bibfnamefont {Aaron}\ \bibnamefont {Chou}} \emph {et~al.},\ }\bibfield  {title} {\enquote {\bibinfo {title} {{Quantum Sensors for High Energy Physics}},}\ \ }(\bibinfo {year} {2023})\ \Eprint {http://arxiv.org/abs/2311.01930} {arXiv:2311.01930 [hep-ex]} \BibitemShut {NoStop}%
\bibitem [{\citenamefont {Dixit}\ \emph {et~al.}(2021)\citenamefont {Dixit}, \citenamefont {Chakram}, \citenamefont {He}, \citenamefont {Agrawal}, \citenamefont {Naik}, \citenamefont {Schuster},\ and\ \citenamefont {Chou}}]{Dixit:2020ymh}%
  \BibitemOpen
  \bibfield  {author} {\bibinfo {author} {\bibfnamefont {Akash~V.}\ \bibnamefont {Dixit}}, \bibinfo {author} {\bibfnamefont {Srivatsan}\ \bibnamefont {Chakram}}, \bibinfo {author} {\bibfnamefont {Kevin}\ \bibnamefont {He}}, \bibinfo {author} {\bibfnamefont {Ankur}\ \bibnamefont {Agrawal}}, \bibinfo {author} {\bibfnamefont {Ravi~K.}\ \bibnamefont {Naik}}, \bibinfo {author} {\bibfnamefont {David~I.}\ \bibnamefont {Schuster}}, \ and\ \bibinfo {author} {\bibfnamefont {Aaron}\ \bibnamefont {Chou}},\ }\bibfield  {title} {\enquote {\bibinfo {title} {{Searching for Dark Matter with a Superconducting Qubit}},}\ }\href {\doibase 10.1103/PhysRevLett.126.141302} {\bibfield  {journal} {\bibinfo  {journal} {Phys. Rev. Lett.}\ }\textbf {\bibinfo {volume} {126}},\ \bibinfo {pages} {141302} (\bibinfo {year} {2021})},\ \Eprint {http://arxiv.org/abs/2008.12231} {arXiv:2008.12231 [hep-ex]} \BibitemShut {NoStop}%
\bibitem [{\citenamefont {Chen}\ \emph {et~al.}(2023)\citenamefont {Chen}, \citenamefont {Fukuda}, \citenamefont {Inada}, \citenamefont {Moroi}, \citenamefont {Nitta},\ and\ \citenamefont {Sichanugrist}}]{Chen:2022quj}%
  \BibitemOpen
  \bibfield  {author} {\bibinfo {author} {\bibfnamefont {Shion}\ \bibnamefont {Chen}}, \bibinfo {author} {\bibfnamefont {Hajime}\ \bibnamefont {Fukuda}}, \bibinfo {author} {\bibfnamefont {Toshiaki}\ \bibnamefont {Inada}}, \bibinfo {author} {\bibfnamefont {Takeo}\ \bibnamefont {Moroi}}, \bibinfo {author} {\bibfnamefont {Tatsumi}\ \bibnamefont {Nitta}}, \ and\ \bibinfo {author} {\bibfnamefont {Thanaporn}\ \bibnamefont {Sichanugrist}},\ }\bibfield  {title} {\enquote {\bibinfo {title} {{Detecting Hidden Photon Dark Matter Using the Direct Excitation of Transmon Qubits}},}\ }\href {\doibase 10.1103/PhysRevLett.131.211001} {\bibfield  {journal} {\bibinfo  {journal} {Phys. Rev. Lett.}\ }\textbf {\bibinfo {volume} {131}},\ \bibinfo {pages} {211001} (\bibinfo {year} {2023})},\ \Eprint {http://arxiv.org/abs/2212.03884} {arXiv:2212.03884 [hep-ph]} \BibitemShut {NoStop}%
\bibitem [{\citenamefont {Chen}\ \emph {et~al.}(2024{\natexlab{a}})\citenamefont {Chen}, \citenamefont {Fukuda}, \citenamefont {Inada}, \citenamefont {Moroi}, \citenamefont {Nitta},\ and\ \citenamefont {Sichanugrist}}]{Chen:2023swh}%
  \BibitemOpen
  \bibfield  {author} {\bibinfo {author} {\bibfnamefont {Shion}\ \bibnamefont {Chen}}, \bibinfo {author} {\bibfnamefont {Hajime}\ \bibnamefont {Fukuda}}, \bibinfo {author} {\bibfnamefont {Toshiaki}\ \bibnamefont {Inada}}, \bibinfo {author} {\bibfnamefont {Takeo}\ \bibnamefont {Moroi}}, \bibinfo {author} {\bibfnamefont {Tatsumi}\ \bibnamefont {Nitta}}, \ and\ \bibinfo {author} {\bibfnamefont {Thanaporn}\ \bibnamefont {Sichanugrist}},\ }\bibfield  {title} {\enquote {\bibinfo {title} {{Quantum Enhancement in Dark Matter Detection with Quantum Computation}},}\ }\href {\doibase 10.1103/PhysRevLett.133.021801} {\bibfield  {journal} {\bibinfo  {journal} {Phys. Rev. Lett.}\ }\textbf {\bibinfo {volume} {133}},\ \bibinfo {pages} {021801} (\bibinfo {year} {2024}{\natexlab{a}})},\ \Eprint {http://arxiv.org/abs/2311.10413} {arXiv:2311.10413 [hep-ph]} \BibitemShut {NoStop}%
\bibitem [{\citenamefont {Chen}\ \emph {et~al.}(2024{\natexlab{b}})\citenamefont {Chen}, \citenamefont {Fukuda}, \citenamefont {Inada}, \citenamefont {Moroi}, \citenamefont {Nitta},\ and\ \citenamefont {Sichanugrist}}]{Chen:2024aya}%
  \BibitemOpen
  \bibfield  {author} {\bibinfo {author} {\bibfnamefont {Shion}\ \bibnamefont {Chen}}, \bibinfo {author} {\bibfnamefont {Hajime}\ \bibnamefont {Fukuda}}, \bibinfo {author} {\bibfnamefont {Toshiaki}\ \bibnamefont {Inada}}, \bibinfo {author} {\bibfnamefont {Takeo}\ \bibnamefont {Moroi}}, \bibinfo {author} {\bibfnamefont {Tatsumi}\ \bibnamefont {Nitta}}, \ and\ \bibinfo {author} {\bibfnamefont {Thanaporn}\ \bibnamefont {Sichanugrist}},\ }\bibfield  {title} {\enquote {\bibinfo {title} {{Search for QCD axion dark matter with transmon qubits and quantum circuit}},}\ }\href@noop {} {\  (\bibinfo {year} {2024}{\natexlab{b}})},\ \Eprint {http://arxiv.org/abs/2407.19755} {arXiv:2407.19755 [hep-ph]} \BibitemShut {NoStop}%
\bibitem [{\citenamefont {Chigusa}\ \emph {et~al.}(2023)\citenamefont {Chigusa}, \citenamefont {Hazumi}, \citenamefont {Herbschleb}, \citenamefont {Mizuochi},\ and\ \citenamefont {Nakayama}}]{Chigusa:2023hms}%
  \BibitemOpen
  \bibfield  {author} {\bibinfo {author} {\bibfnamefont {So}~\bibnamefont {Chigusa}}, \bibinfo {author} {\bibfnamefont {Masashi}\ \bibnamefont {Hazumi}}, \bibinfo {author} {\bibfnamefont {Ernst~David}\ \bibnamefont {Herbschleb}}, \bibinfo {author} {\bibfnamefont {Norikazu}\ \bibnamefont {Mizuochi}}, \ and\ \bibinfo {author} {\bibfnamefont {Kazunori}\ \bibnamefont {Nakayama}},\ }\bibfield  {title} {\enquote {\bibinfo {title} {{Light Dark Matter Search with Nitrogen-Vacancy Centers in Diamonds}},}\ }\href@noop {} {\  (\bibinfo {year} {2023})},\ \Eprint {http://arxiv.org/abs/2302.12756} {arXiv:2302.12756 [hep-ph]} \BibitemShut {NoStop}%
\bibitem [{\citenamefont {Chigusa}\ \emph {et~al.}(2024)\citenamefont {Chigusa}, \citenamefont {Hazumi}, \citenamefont {Herbschleb}, \citenamefont {Matsuzaki}, \citenamefont {Mizuochi},\ and\ \citenamefont {Nakayama}}]{Chigusa:2024psk}%
  \BibitemOpen
  \bibfield  {author} {\bibinfo {author} {\bibfnamefont {So}~\bibnamefont {Chigusa}}, \bibinfo {author} {\bibfnamefont {Masashi}\ \bibnamefont {Hazumi}}, \bibinfo {author} {\bibfnamefont {Ernst~David}\ \bibnamefont {Herbschleb}}, \bibinfo {author} {\bibfnamefont {Yuichiro}\ \bibnamefont {Matsuzaki}}, \bibinfo {author} {\bibfnamefont {Norikazu}\ \bibnamefont {Mizuochi}}, \ and\ \bibinfo {author} {\bibfnamefont {Kazunori}\ \bibnamefont {Nakayama}},\ }\bibfield  {title} {\enquote {\bibinfo {title} {{Nuclear Spin Metrology with Nitrogen Vacancy Center in Diamond for Axion Dark Matter Detection}},}\ }\href@noop {} {\  (\bibinfo {year} {2024})},\ \Eprint {http://arxiv.org/abs/2407.07141} {arXiv:2407.07141 [hep-ph]} \BibitemShut {NoStop}%
\bibitem [{\citenamefont {Ito}\ \emph {et~al.}(2024)\citenamefont {Ito}, \citenamefont {Kitano}, \citenamefont {Nakano},\ and\ \citenamefont {Takai}}]{Ito:2023zhp}%
  \BibitemOpen
  \bibfield  {author} {\bibinfo {author} {\bibfnamefont {Asuka}\ \bibnamefont {Ito}}, \bibinfo {author} {\bibfnamefont {Ryuichiro}\ \bibnamefont {Kitano}}, \bibinfo {author} {\bibfnamefont {Wakutaka}\ \bibnamefont {Nakano}}, \ and\ \bibinfo {author} {\bibfnamefont {Ryoto}\ \bibnamefont {Takai}},\ }\bibfield  {title} {\enquote {\bibinfo {title} {{Quantum entanglement of ions for light dark matter detection}},}\ }\href {\doibase 10.1007/JHEP02(2024)124} {\bibfield  {journal} {\bibinfo  {journal} {JHEP}\ }\textbf {\bibinfo {volume} {02}},\ \bibinfo {pages} {124} (\bibinfo {year} {2024})},\ \Eprint {http://arxiv.org/abs/2311.11632} {arXiv:2311.11632 [hep-ph]} \BibitemShut {NoStop}%
\bibitem [{\citenamefont {Ballmer}\ \emph {et~al.}(2022)\citenamefont {Ballmer} \emph {et~al.}}]{Ballmer:2022uxx}%
  \BibitemOpen
  \bibfield  {author} {\bibinfo {author} {\bibfnamefont {Stefan~W.}\ \bibnamefont {Ballmer}} \emph {et~al.},\ }\bibfield  {title} {\enquote {\bibinfo {title} {{Snowmass2021 Cosmic Frontier White Paper: Future Gravitational-Wave Detector Facilities}},}\ }in\ \href@noop {} {\emph {\bibinfo {booktitle} {{Snowmass 2021}}}}\ (\bibinfo {year} {2022})\ \Eprint {http://arxiv.org/abs/2203.08228} {arXiv:2203.08228 [gr-qc]} \BibitemShut {NoStop}%
\bibitem [{\citenamefont {Ito}\ and\ \citenamefont {Kitano}(2024)}]{Ito:2023bnu}%
  \BibitemOpen
  \bibfield  {author} {\bibinfo {author} {\bibfnamefont {Asuka}\ \bibnamefont {Ito}}\ and\ \bibinfo {author} {\bibfnamefont {Ryuichiro}\ \bibnamefont {Kitano}},\ }\bibfield  {title} {\enquote {\bibinfo {title} {{Macroscopic quantum response to gravitational waves}},}\ }\href {\doibase 10.1088/1475-7516/2024/04/068} {\bibfield  {journal} {\bibinfo  {journal} {JCAP}\ }\textbf {\bibinfo {volume} {04}},\ \bibinfo {pages} {068} (\bibinfo {year} {2024})},\ \Eprint {http://arxiv.org/abs/2309.02992} {arXiv:2309.02992 [gr-qc]} \BibitemShut {NoStop}%
\end{thebibliography}%

\end{document}